\documentclass[a4paper,fleqn]{cas-dc}

\usepackage{graphicx}
\usepackage[font=scriptsize, justification=centering]{caption}
\usepackage{subcaption}
\usepackage[switch]{lineno}
\usepackage{multirow}
\usepackage{enumitem}
\usepackage{listings}
\usepackage{xspace}
\usepackage[flushleft]{threeparttable}
\usepackage[numbers]{natbib}
\usepackage{color,soul}

\def\tsc#1{\csdef{#1}{\textsc{\lowercase{#1}}\xspace}}
\tsc{WGM}
\tsc{QE}
\tsc{EP}
\tsc{PMS}
\tsc{BEC}
\tsc{DE}

\newdefinition{definition}{Definition}
\newcommand{\etal}{\textit{et al.}}

\definecolor{ao}{rgb}{0.0, 0.5, 0.0}
\newcommand{\tool}{\textsc{ARist}\xspace}
\newcommand{\gpt}{GPT-2\xspace}

\begin{document}
\let\WriteBookmarks\relax
\def\floatpagepagefraction{1}
\def\textpagefraction{.001}

\shorttitle{\tool}

\shortauthors{Nguyen et~al.}

\title [mode = title]{\tool: An Effective API Argument Recommendation Approach}    



%

\author{Son Nguyen}[orcid=0000-0002-8970-9870]
\ead{sonnguyen@vnu.edu.vn}



\affiliation{organization={Faculty of Information Technology, VNU University of Engineering and Technology},
    city={Hanoi},
    country={Vietnam}}





\author{Cuong Tran Manh}
\ead{tranmanhcuong@vnu.edu.vn}
\author{Kien T. Tran}
\ead{18020026@vnu.edu.vn}
\author{Tan M. Nguyen}
\ead{18020050@vnu.edu.vn}

\author{Thu-Trang Nguyen}[orcid=0000-0002-3596-2352]
\ead{trang.nguyen@vnu.edu.vn}

\author{Kien-Tuan Ngo}[orcid=0000-0001-7136-7529]
\ead{tuanngokien@vnu.edu.vn}

\author{Hieu Dinh Vo}[orcid=0000-0002-9407-1971]
\ead{hieuvd@vnu.edu.vn}
\cormark[1]






\cortext[cor1]{Corresponding author}




\begin{abstract}
Learning and remembering to use APIs are difficult. Several techniques have been proposed to assist developers in using APIs. Most existing techniques focus on recommending the right API methods to call, but very few techniques focus on recommending API arguments. 
In this paper, we propose \tool, a novel automated argument recommendation approach which suggests arguments by predicting developers' \textit{expectations} when they define and use API methods. To implement this idea in the recommendation process, \tool combines program analysis (PA), language models (LMs), and several features specialized for the recommendation task which consider the functionality of formal parameters and the positional information of code elements (e.g., variables or method calls) in the given context. 
In \tool, the LMs and the recommending features are used to suggest the promising candidates identified by PA. Meanwhile, PA navigates the LMs and the features working on the set of the valid candidates which satisfy syntax, accessibility, and type-compatibility constraints defined by the programming language in use. 
Our evaluation on a large dataset of real-world projects shows that \tool improves the state-of-the-art approach by 19\% and 18\% in top-1 precision and recall for recommending arguments of frequently-used libraries. For general argument recommendation task, i.e., recommending arguments for every method call, \tool outperforms the baseline approaches by up to 125\% top-1 accuracy. Moreover, for newly-encountered projects, \tool achieves more than 60\% top-3 accuracy when evaluating on a larger dataset. For working/maintaining projects, with a personalized LM to capture developers' coding practice, \tool can productively rank the expected arguments at the top-1 position in 7/10 requests.
\end{abstract}




\maketitle

\section{Introduction}
\label{sec:introduction}
Application programming interfaces (APIs) are extensively used in software development to reuse defined components in source code such as libraries and frameworks. Learning and remembering to use APIs are difficult.
Thus, researchers have introduced the techniques to help developers in using APIs~\cite{nguyen2016api,chen2021api,huang2018api,he2021pyart,PARC,Precise,lexsimpaper}.
Most existing approaches focus on recommending the right methods to call and leave the arguments for developers to complete~\cite{nguyen2016api,chen2021api,huang2018api,he2021pyart}.
According to the empirical studies in the existing work~\cite{Precise}, more than haft of method calls in practice require arguments (aka., actual parameters), and unfamiliarity with argument usage could cause bugs~\cite{argument_defects, Precise,pradel2012static}.
Moreover, effectively determining the correct arguments passed to method calls is a non-trivial task~\cite{PARC,Precise,lexsimpaper}. 
Indeed, the arguments usually contain one or more identifiers, while recommending identifiers remains one of the most challenging tasks in automated code completion~\cite{bigcode,curglm,failures}. 
%


Recognizing the importance of argument recommendation, researchers have introduced automated tools to suggest arguments to complete method calls. 
Asaduzzaman \etal~\cite{PARC} and Zhang \etal~\cite{Precise} follow an information retrieval (IR) direction to recommend arguments for the APIs that have been frequently used in the past. These approaches construct a database of the seen arguments and their corresponding context. For a request from developers to recommend an argument, hereinafter \textit{argument recommendation request}, the key idea is that two similar contexts should have similar arguments.
The IR direction requires an appropriate design of predefined static features to represent contexts and methods determining their similarity. For example, the coding context could be represented by containing (enclosing) method name/declaration, calling methods, or the code tokens prior to the calling methods~\cite{PARC,Precise}. 
More importantly, since the argument candidates are suggested by searching for the similar contexts, these IR approaches cannot recommend a new argument that they have not seen before.
%

%
Our empirical study performed on 1,000 top-ranked Java projects on Github (Section~\ref{sec:example}) shows that arguments and their usages (an argument with an API method call) are very unique. Specifically, 63\% of the argument usages are unique, while the proportion of unique arguments is 47\%. 
%
%
Hence, searching for the previously seen arguments to complete API method calls may not be effective.
Moreover, the recommended arguments must conform to the syntactic, type-comparability, and accessibility constraints defined by programming languages. Thus, using the seen arguments without considering those constrains might cause incorrect recommendations.


In this work, we introduce \tool, a novel approach which combines program analysis (PA), language models (LMs), and several argument-recommending features considering the functionality of formal parameters and positional information to predict developers' expected arguments. For an argument recommendation request of an API method call, our idea is that the argument candidates must conform to the syntactic and semantic constraints defined by the programming language in use. Also, the candidates should meet the expectations of the API developers and API users. We aim to benefit from the strengths of both PA and LMs as well as the recommending features in which the LMs and the features are applied to suggest the arguments that satisfy the stakeholders' expectations, while PA enforces the syntactic, accessibility, and type-compatibility constraints in the recommendation process. 

Particularly, \tool applies PA to analyze the given partially completed code (\textit{context}) to identify the syntax-valid, accessible, and type-compatible argument candidates (so-called \textit{valid candidates}) rather than retrieving candidates from a set of the seen arguments. 
The identified valid candidates are ranked according to their likelihood to be the next argument of the API method call. In \tool, the likelihood of candidates is assigned by the LMs and the designed features. 
%
%
%
However, the identified set of valid candidates could be huge. The reason is that the accessible code elements (e.g., variables, fields, methods, or literals) in source code could be combined in many ways to form type-compatible candidates (i.e., expressions) for the request. 
This problem of a huge valid candidate set could cause the ineffectiveness and inefficiency of the ranking stage. To tackle this problem, before ranking the identified valid candidates, we filter out the less promising candidates by applying a set of heuristic rules and a light-ranking stage. 

%
%
%

To evaluate \tool, we conducted several experiments on two large datasets used in the existing studies with +1.6M argument recommendation requests in two very large projects (Eclipse and Netbeans)~\cite{PARC} and 9.2K projects~\cite{bigcode}, respectively. 
Our results show that \tool outperforms the state-of-the-art approach~\cite{PARC} in both precision and recall up to 19\% and 18\% in recommending arguments for \textit{frequently-used libraries} of Eclipse and Netbeans. For \textit{general argument recommendation} task (i.e., recommending arguments for every method call) in these projects, the performance of \tool is also significantly better than that of the state-of-the-art $n$-gram LM with nested cache (SLP)~\cite{nestedcache}, GPT-2~\cite{gpt2}, and CodeT5~\cite{codet5} which are selected as the representative baseline approaches in general argument recommendation. Notably, GPT-2 is the core engine of Tabnine~\cite{tabnine} and IntelliCode~\cite{intellicode}, two of the most popular AI-assisted code completer. Particularly, \tool improves up to 125\% in top-1 accuracy for CodeT5, GPT-2, and SLP.
Moreover, after training with the much larger dataset~\cite{bigcode}, \tool achieves more than 60\% top-3 accuracy in recommending arguments for \textit{newly-encountered projects}. 
%
%
For \textit{working/maintaining projects}, with a personalized LM to capture developers’ coding practice, \tool can productively rank the expected arguments at the top-1 position in up to +7/10 recommendation requests. 

In brief, this paper makes the following contributions:
\begin{enumerate}
    \item \tool, a novel argument recommendation tool combining program analysis and language models in an effective and efficient recommendation process.
    \item An extensive empirical evaluation showing that \tool significantly outperforms the baseline approaches, and \tool is effective in assisting developers in coding tasks.
\end{enumerate}

\section{Empirical Study}
\label{sec:example}
For automatically recommending arguments, one would face the following challenges. First, the recommended candidates must conform to the syntactic, type-compatibility, and accessibility constraints defined by the programming language in use. Second, the tool must correctly predict the argument which a developer intends to fill the method call at hand. To better understand the characteristics of arguments which might affect the automation of argument recommending, we conducted an empirical study to answer the following research questions:




\textbf{RQ1:}\textit{ What are the expression types~\cite{java_spec} of arguments?} For this RQ, we explore how common expression types (e.g., Simple Name, Method Invocation or String Literal) of arguments are in practice. 

\textbf{RQ2:} \textit{What are the expected data types of arguments?} For this RQ, we investigate how common the data types (e.g., \texttt{String} or \texttt{Object}) of arguments are expected by the calling methods. In attempt to estimate the numbers  of valid candidates for each expected data type of arguments, we design a simple method to generate syntax-valid, accessible, and type-valid candidates for the calling methods.

\textbf{RQ3:} \textit{How unique are the arguments and their usages (i.e., an argument with a calling method)?} For this RQ, we investigate how often an argument of a calling method is likely to be not seem before (i.e., unique).

\textbf{RQ4:} \textit{Where are the arguments from?} For this RQ, we investigate how often an argument is in the same projects as their calling methods.

\begin{definition}
An \textit{argument usage} is a 3-tuple $\langle arg, m\_call, pos \rangle$ where $arg$ is an argument name, $m\_call$ is a calling method name in which $arg$ is the $pos^{th}$ argument.
\end{definition}

\begin{definition}
An \textit{Argument Recommendation (AR) request} $r$ is a 3-tuple $r=\langle P, m\_call, pos \rangle$ where $P$ is a partially completed code (context), $m\_call$ is a calling method which is partially completed in $P$ where the $pos^{th}$ argument of $m\_call$ is unfilled and requested for recommendations.
\end{definition}

\subsection{Data collection and processing} 

In this empirical study, we collected a dataset of 1,000 top-ranked, high-quality, long-history Java projects on GitHub. The projects are ranked based on the numbers of stars, folks, and commits of the repositories. The dataset could be found on our website~\cite{ourwebsite}.
In this dataset, all duplicated Java files, migrated projects, and the forks of the same projects are filtered out. This dataset includes about 460K files, +4.2M methods, and +42.4M LOC. 
For each method call having arguments, we collected and analyzed the name and origin of the calling method as well as the names, expression types, expected data types, and origin of arguments. 


%

\subsection{Results and Analysis}

\begin{figure*}
\centering
\begin{subfigure}{.5\textwidth}
  \centering
  \includegraphics[width=1\linewidth]{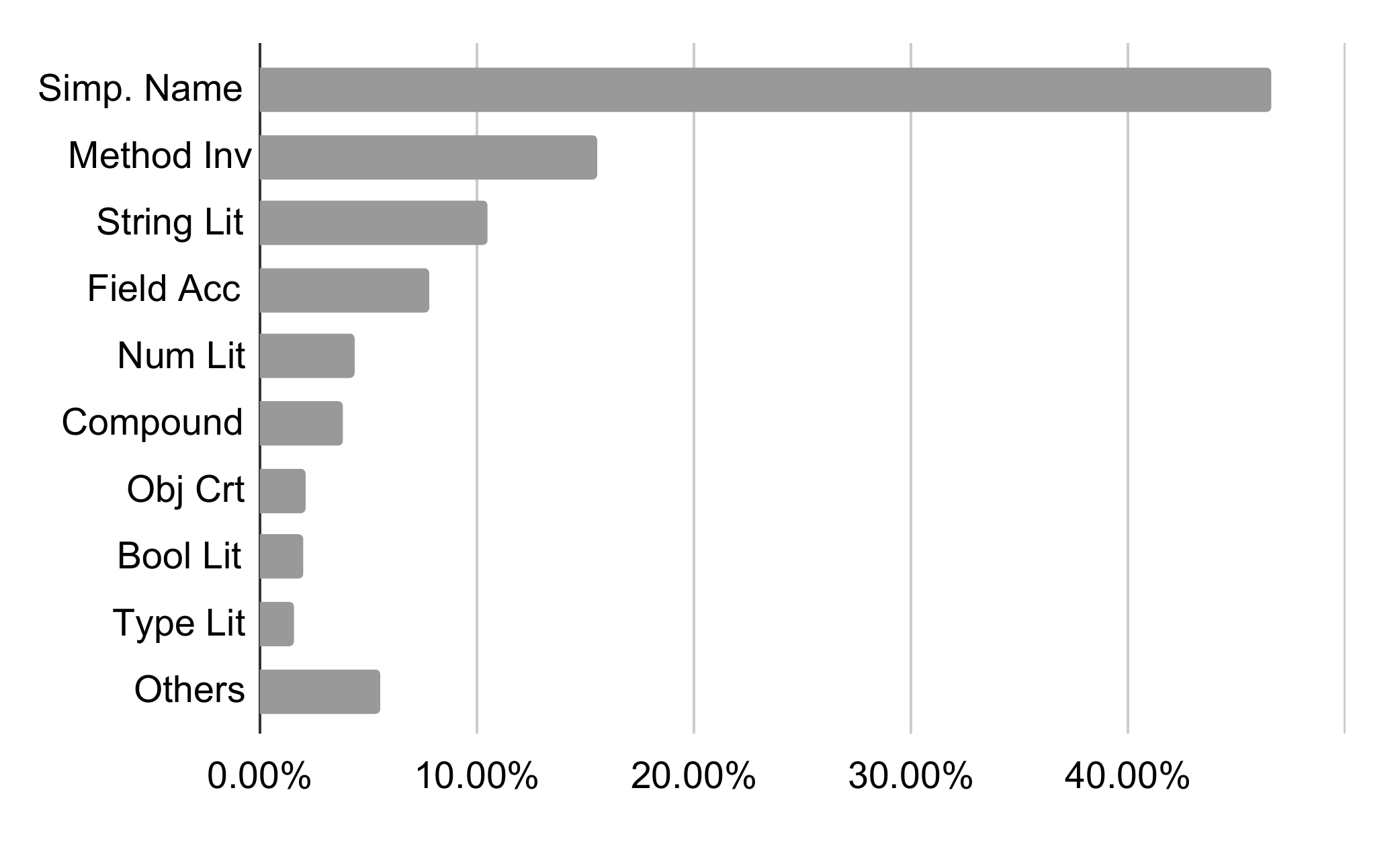}
  \caption{}
  \label{fig:frequency_exp_type}
\end{subfigure}%
\begin{subfigure}{.5\textwidth}
  \centering
  \includegraphics[width=1\linewidth]{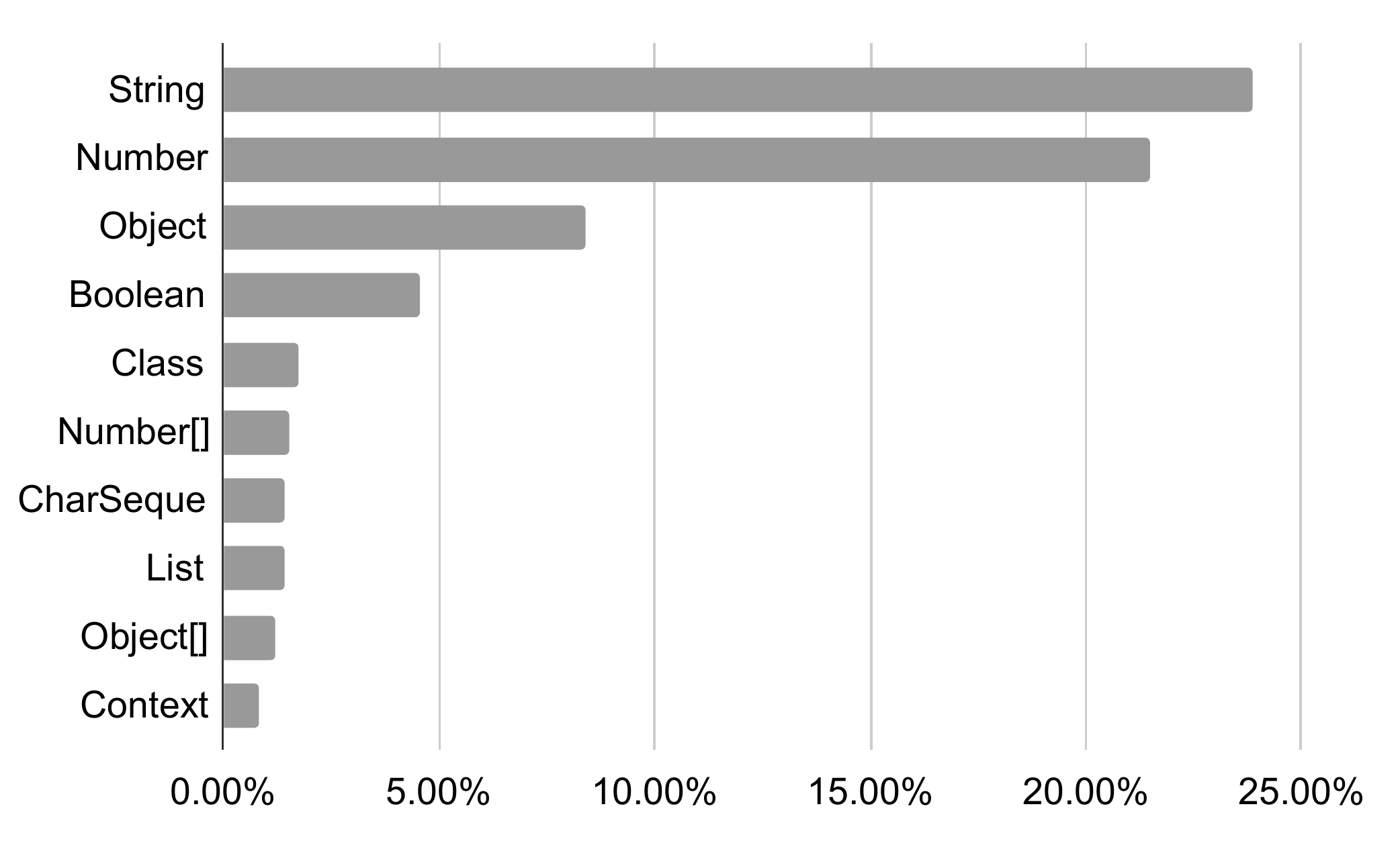}
  \caption{}
  \label{fig:frequency_data_type}
\end{subfigure}

\caption{(a): Argument usage distribution by expr. types~\cite{java_spec}; (b): Top-10 most popular argument data types}
\label{fig:test}
\end{figure*}

\textit{RQ1:}\textit{ What are the expression types~\cite{java_spec} of arguments?} Figure~\ref{fig:frequency_exp_type} shows the portions of argument usages by expression types~\cite{jdt_specification,java_spec, Precise}.
As seen, 70\% of arguments are \textit{Simple Name}, \textit{Method Invocation}, and \textit{Field Access}. In other words, \textit{at least seven out of ten arguments are/contain identifiers}.  
Meanwhile, existing studies~\cite{bigcode,curglm} have confirmed that recommending identifiers remains a much harder task compared to recommending other token kinds such as keywords, separators, and operators of the syntactic grammar~\cite{javagrammar}.
We also found that 60\% of method calls have at least one argument in our dataset. Thus, \textit{argument recommendation is a very frequent task, still, AR is very hard.}

%

\textit{RQ2:} \textit{What are the expected data types of arguments?} As seen in Figure~\ref{fig:frequency_data_type}, a significant portion of argument usages require an argument in the common data types~\cite{jdt_specification,java_spec}, such as \texttt{Object} or \texttt{String}. 
About 54\% of the usages expect that the data type of arguments which is \texttt{Object}, \texttt{String}, or Number (e.g., \texttt{int} or \texttt{Double}). 
Consequently, when argument recommendations are requested at such positions, there are very large sets of (accessible) argument candidates that are type-compatible. 
Indeed, there are more than a half of argument usages (about 550K/1M usages) having at least 100 type-compatible (\textit{valid}) candidates. 
Note that the valid candidates are limited by the set of expression types in Table \ref{tab:expr_types}. Without limiting expression types, the number of valid candidates could be infinite. 
%
For example, there are many argument usages in Eclipse project where the number of valid alternatives is up to 100K.
Thus, given an AR request, \textit{the set of argument candidates which are valid could be very large, and identifying the expected argument could be a challenging task}.

\begin{figure}
\centering
\begin{subfigure}{.25\textwidth}
  \centering
  \includegraphics[width=1\linewidth]{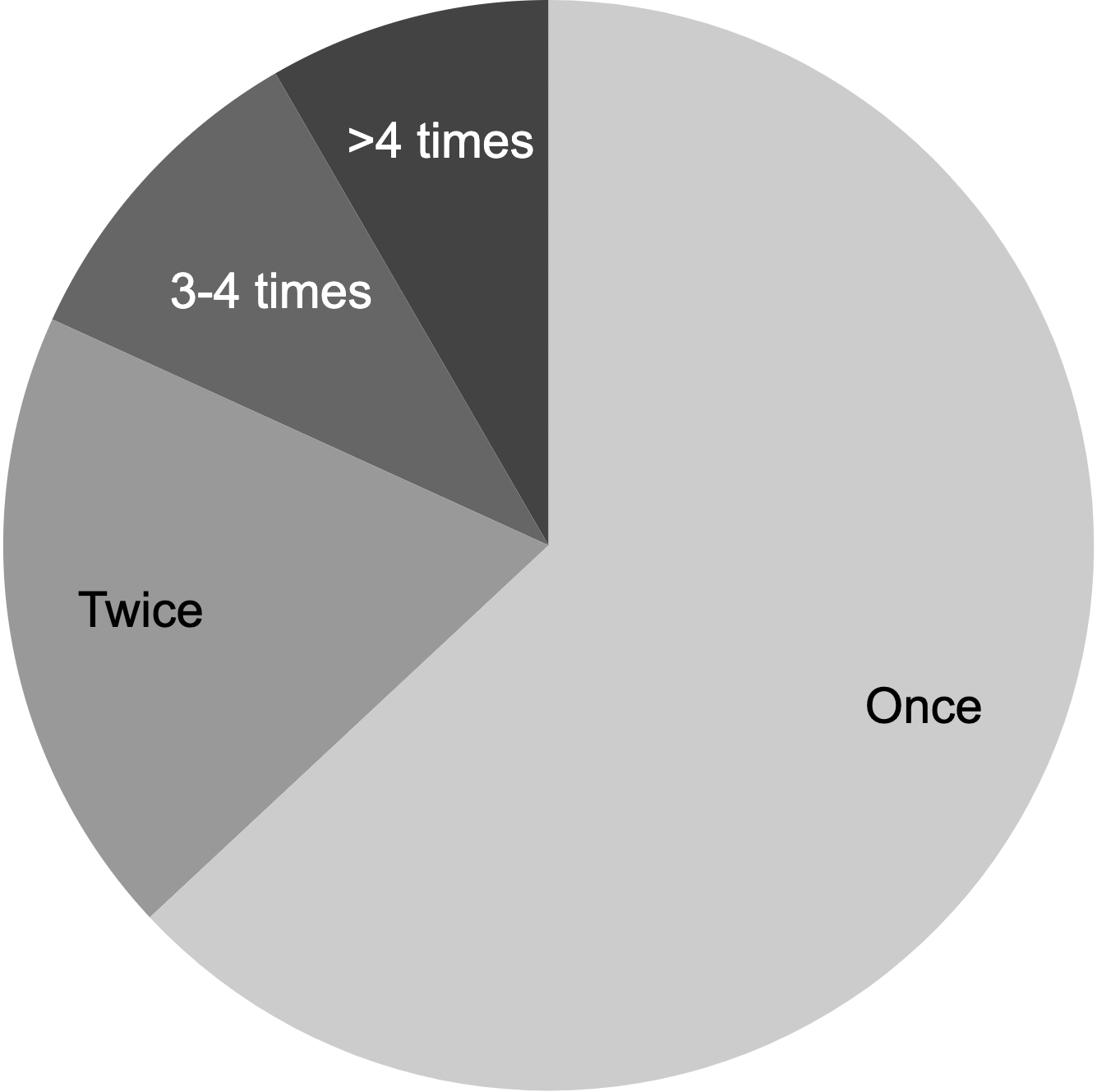}
  \caption{}
  \label{fig:frequency_arg_usage}
\end{subfigure}%
%
%
\begin{subfigure}{.25\textwidth}
  \centering
  \includegraphics[width=1\linewidth]{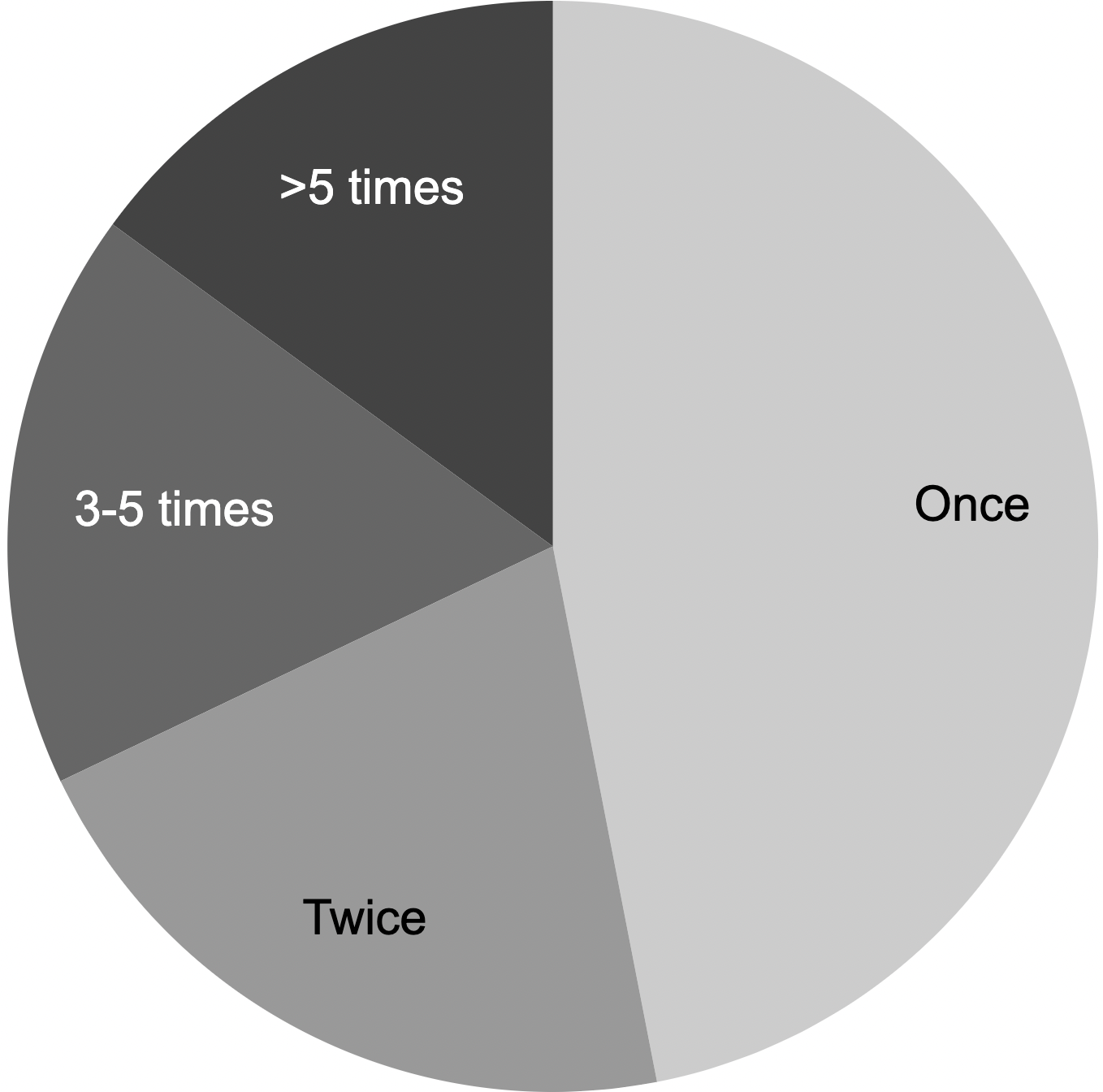}
  \caption{}
  \label{fig:argument_freq}
\end{subfigure}

\caption{(a): Argument usage frequency; (b): Sole argument frequency}
\end{figure}

\textit{RQ3:} \textit{How unique are the arguments and their usages (i.e., an argument with a calling method)?}       
%
In our large corpus, argument usages are quite unique. 
Indeed, there are more than 6 out of 10 usages which appear \textbf{only once} in our dataset (Figure~\ref{fig:frequency_arg_usage}). 
Even for sole arguments (the argument part in argument usages), they are still quite infrequent: 47\% of the arguments are unique (Figure~\ref{fig:argument_freq}). 
As a result, when being requested for recommendations, \textit{a very large number of the arguments might not be identified by searching for similar arguments/usages in the set of cases that are previously encountered}.

\textit{RQ4:} \textit{Where are the arguments from?}
Regarding the origin of calling methods and arguments, nearly 50\% of method calls are the invocations of the methods in the containing projects, while 84\% of arguments can be only found in the containing projects. For these cases, those statistical approaches might fail to learn argument usages from a set of projects to apply to other projects. Moreover, the methods having the same method name can be declared with different sets of parameters. For example, there are several methods named \texttt{insert}, yet declared with different sets of parameters, such as \texttt{int insert(User u)}, \texttt{void insert(Record rec)}, or \texttt{void insert(int off, char[] str)}. 
In these cases,\textit{ one might not have a correct suggestion by blindly using the examples from the seen projects to recommend arguments in another project}.

    
    
    

\section{Key Ideas and Approach Design}
\label{sec:idea_approach}

\subsection{Key Ideas}
To address the challenges discussed above, in completing argument for an API method, $m\_call$, our idea is to recommend the argument candidates which satisfy the expectation of three stakeholders: the \textit{creator} of the programming language in use, \textit{API developers}, and \textit{API users}. 

For the \textit{language creators}, the argument candidates are expected to be \textit{valid} in terms of the syntax, type-compatibility, and accessibility. These expectations are encoded in the corresponding language compiler/interpreter as a set of rules to ensure that the written programs are valid/well-form. 
For the argument recommendation (AR) request in Figure~\ref{fig:example_filtering}, the argument candidates must be syntax-valid~\cite{java_spec} and type-comparable expressions such as a variable (Simple Name) typed \texttt{Object} or a \texttt{String} literal. Additionally, the candidates must be accessible from the requesting location. In Figure~\ref{fig:example_filtering}, a variable candidate must be either fields of \texttt{AnnotationTypes} or static variables in Nebeans.
Thus, in the argument recommendation process, the syntax, type-compatibility, and accessibility rules should be enforced to determine whether an argument candidate is \textit{valid} or not.

For \textit{API developers}, when defining $m\_call$, they might expect that the future arguments of $m\_call$ must be \textit{compatible} with the corresponding parameters in terms of their datatype and functionality. 
The type-compatibility expected by the language creators is \textit{general} for all types developed in the language. Meanwhile, the type-compatibility expectation of the API developer is \textit{specific} for the type of the corresponding parameter. Additionally, the functionality of parameters and arguments could be carried by their name. 
For example, the argument of \texttt{getProp} defined at lines 5--7 of Figure~\ref{fig:example_filtering}, when the method is called at line 10, is expected to be a \textbf{prop}erty. The correct argument is \texttt{PROP\_BACKGROUND\_DRAWING} which indeed is a property.
%
%
Liu \etal~\cite{lexsimpaper} show that arguments are very similar or even identical to their corresponding parameters. Inspired by this study, we estimate how the candidate fulfills the developers' expectation by measuring the similarity between the candidate's name and the corresponding parameter's name of $m\_call$ (\textit{parameter similarity}, Section \ref{sec:para_sim}). 

The \textit{API users} also expect that the candidates are valid regarding the programming language and reflect the users' intention. 
%
Indeed, the specific given context with each candidate must conform to the rules about the syntax, type-compatibility, and accessibility of the language in use.
Additionally, to predict the intention of the developers when they use $m\_call$, we rely on the naturalness (repetition) of code because source code naturally written by developers does not occur randomly~\cite{naturalness}. 
For example, the candidates of the AR request inside \texttt{isBackgroundDrawing} should intuitively be a property related to \texttt{BackgroundDrawing}. Thus, compared to \texttt{PROP\_COMBINE\_GLYPHS}, \texttt{PROP\_BACKGROUND\_DRAWING} is more likely to be the expected argument of the AR request in Figure~\ref{fig:example_filtering}, 
although both of the candidates are valid in terms of the syntax, type-compatibility, and accessibility.
In fact, the naturalness of code can be captured by language models trained on the seen source code~\cite{naturalness,nestedcache}. 

Additionally, like natural language processing, besides the repetition of code, the positional information, which gives the model the relative position of the tokens in the context, also plays an important role in predicting the next tokens/argument~\cite{vaswani2017attention}. Inspired by this idea, to capture the positional information of the candidates and apply it to predict the next arguments, we design two features specialized for the recommendation task, \textit{creating-distance} and \textit{accessing-recentness}, which are the ``distances'' to the positions where the candidate is declared and defined/used (Section \ref{sec:recentness}).

\begin{figure}
\centering
\begin{lstlisting}[language=Java]
package org.netbeans.editor;
public class AnnotationTypes {
  public static String PROP_BACKGROUND_DRAWING = ...;
  public static String PROP_COMBINE_GLYPHS = ...;
  public Object getProp(Object prop){
    //...
  }
  public Boolean isBackgroundDrawing() {
     loadSettings();
     bool b = (bool) getProp(/*AR_REQUEST*/);
            //Expected: PROP_BACKGROUND_DRAWING
  }
}
\end{lstlisting}
\caption{An example in Netbeans project}
\label{fig:example_filtering}
\end{figure}

\subsection{Approach Design}

To implement the above ideas in \tool, we combine program analysis (PA), language models (LMs), and several features specialized for the argument recommendation task. 
For an AR request, the syntax, type-compatibility, and accessibility rules are enforced into the recommendation process to identify the \textit{valid} argument candidates. The identified valid candidates are ranked according to their likelihood to be the next argument of the method call. In \tool, the likelihood is assigned by the LMs and the selected features.
In this novel combination, the LMs and selected features are applied to reduce the uncertainty in deciding the most likely candidates in the valid candidate set identified by PA. Meanwhile, PA can be used to navigate the LMs and selected features to apply on the right set of candidates. 
This can also boost the performance of the LMs in recommending the arguments where their usages are less frequent or have not been seen by the models. 
However, the LMs and selected features might still not effectively rank the expected arguments at the top positions because of the large identified candidate sets and the infrequency of arguments. Plus, these large sets could cause inefficiency in candidate ranking, especially for sophisticated LMs. 
To deal with the problem of large sets of valid candidates, \tool reduces the identified candidate sets by filtering out the candidates which are less likely to be the expected arguments. 

In \tool, we design a set of filtering rules to eliminate the unpromising candidates. Additionally, a light-ranking stage is applied using the selected features and a light-weight LM, such as an $n$-gram LM, to efficiently isolate a relatively large number (e.g., 20--100) of promising candidates. Then, these promising candidates are passed to the heavy-ranking stage where a sophisticated model with more expensive computation such as a deep neural network LM is performed to recommend the most appropriate arguments.

However, applying LMs in \tool might face the problem of rare/unseen method calls/arguments~\cite{bigcode} (out of vocabulary). To mitigate this problem, we tokenize source code into sequences of sub-tokens by \texttt{camelCase} and \texttt{under\_score} naming conventions (for the traditional count-based LMs such as $n$-gram) or employing byte pair encoding (BPE)~\cite{bpe} (for more sophisticated models such as GPT-2 or T5).

\section{Features Specialized for API Argument Recommendation}
\label{sec:useful_features}
In \tool, we design a set of \textit{static} features which capture the expectation of developers to estimate the likelihood to be the expected arguments of AR requests.
In this version of \tool, the features specialized for argument recommendation include 
the similarities between the candidates and the corresponding parameter of the calling method (\textit{parameter similarity}) and how recently the candidates are created and accessed in the context (\textit{creating-distance} and \textit{accessing-recentness}).
These features could always be applied to determine candidates' likelihood even when the candidates and calling methods are infrequent or unseen in the past. This could mitigate the negative impact of the out-of-vocabulary problem of LMs in \tool and help it improve the recommendation performance.
%
%

\subsection{Parameter Similarity}
\label{sec:para_sim}

In a method declaration, a parameter's name usually conveys information
about the semantics of the parameter in the method body as well as the meaning of the expected argument when the method is invoked. 
%
Liu \etal~\cite{lexsimpaper} have shown that parameter names and argument names are very similar. 
%
The similarity between an argument ($c$) and its corresponding parameter ($p$) is calculated as follows:
\begin{equation}
\label{parasim}
parasim(c, p) =\frac{|com\_terms(c, p)| + |com\_terms(p, c)|}{|terms(c)| + |terms(p)|}
\end{equation}
In formula (\ref{parasim}), $terms(s)$ is the decomposition of \textit{s}, based on \texttt{camelCase} and \texttt{under\_score} convention. $com\_terms(n_1, n_2)$ returns the longest subsequence of $terms(n_1)$, where each term in the subsequence appears in $terms(n_2)$.

\begin{figure}
    \centering
    \includegraphics[width=\columnwidth]{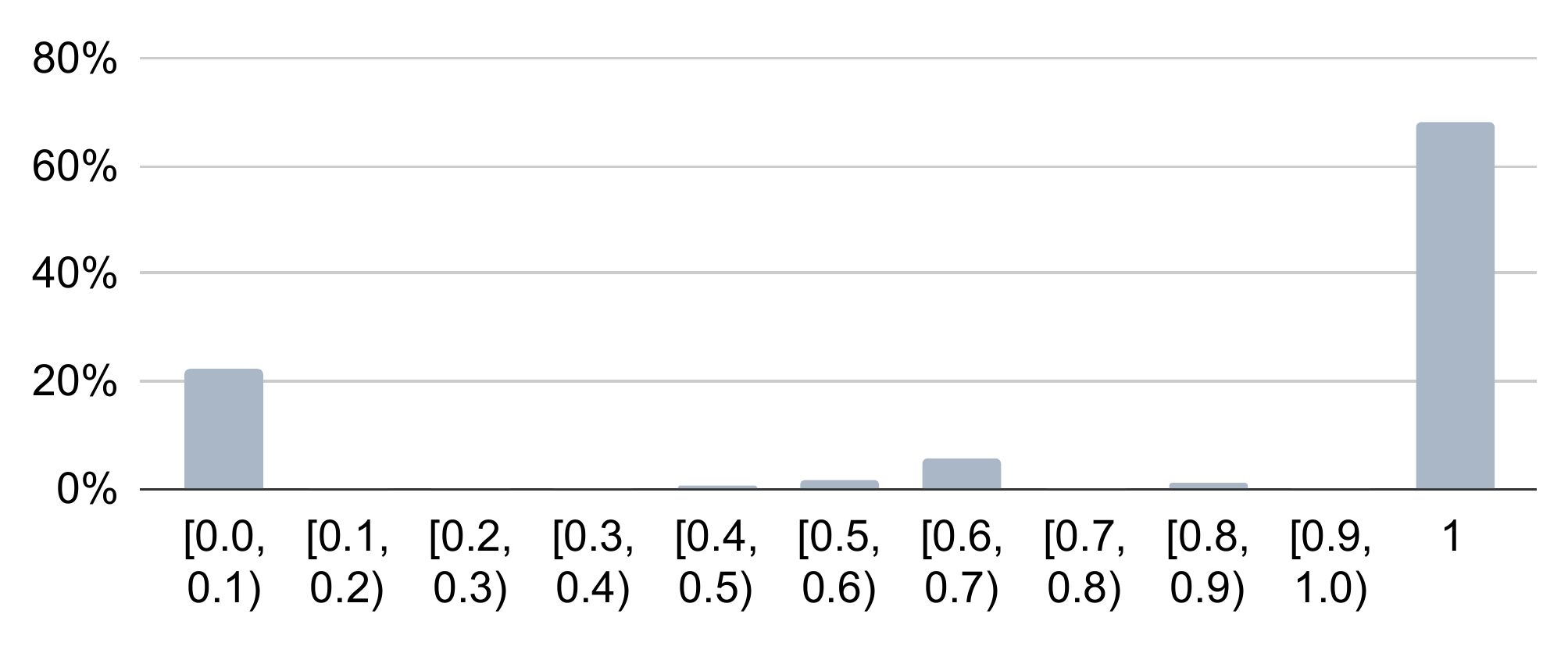}
    \caption{Distribution of parameter similarity}
    \label{fig:parasim}
\end{figure}

We have performed an empirical study to investigate the \textit{parameter similarity} in Eclipse and Netbeans.
Our results show that 68.24\% of the arguments are very similar to their corresponding parameters' names, while 22.33\% of the arguments are completely different (Figure \ref{fig:parasim}). The very low similarity is often caused by short and generic names. This confirmed the characteristic of parameter similarity reported by Liu \etal~\cite{lexsimpaper}. Inspired by Liu \etal~\cite{lexsimpaper}, we also applied parameter similarity to improve the precision of the AR process in \tool.
Note that, to avoid vanishing other features of the candidates which are different from the corresponding parameter, the parameter similarities of the candidates are normalized to the range of $[x,1.0]$ where $x > 0$ to exclude the value of 0.

\subsection{Creating-distance and Accessing-recentness}
\label{sec:recentness}
From a method call's position, the local variables/parameters, instance variables, or class variables (so-called \textit{variables}) used as the arguments of this method call are often \textit{created} or \textit{accessed} very recently. 
In Figure \ref{fig:scope_code}, variable \texttt{provider} is created in the same scope as where it is used as an argument (line 3). 
For the first argument of \texttt{ignoreProvider}, parameter \texttt{user} is also recently \textit{used} at line 2.
In \tool, for a request $r=\langle P, m\_call, pos \rangle$, we utilize the features which reflect how recent the variable is to the first time (\textit{creating-distance}) and the last time (\textit{accessing-recentness}) it appearing in the context to improve the AR performance. 
In practice, these features of a variable might require recording the real-time changes by developers, and hence be hard to collect. 
In \tool, the \textit{creating-distance} and \textit{accessing-recentness} are statically approximated as follows.

\textbf{Creating-distance}. An accessible variable $c$ could be created/declared outside the containing method, even $c$ could be a field of parent classes. 
Thus, the \textit{creating-distance} of $c$ is computed as the \textit{scope-distance} between the code closest block/region containing the calling method, and the code block/region\footnote{A block is a sequence of statements, local class declarations, and local variable declaration statements within braces~\cite{java-grammar}.} where $c$ is created. 

\begin{definition}{\textbf{(creating-distance).}}
\label{dec_r_def}
Given an AR request $r=\langle P, m\_call, pos \rangle$ and a candidate $c$ which is an accessible variable, the creating-distance, $create\_dis(c, r)$, is the scope-distance of $B$ and $B'$, $scope\_dis(B, B')$, where $B$ is the closest block where $c$ is created, and $B'$ is the closest block containing $m\_call$:
\begin{itemize}
    \item $scope\_dis(B, B')=0$ if $B = B'$.
    \item $scope\_dis(B, B')=1+scope\_dis(B, B'')$, where $B''$ is the closest block containing $B'$.
\end{itemize}
\end{definition}

For AR request $r$ in Figure \ref{fig:scope_code}, $B'$ contains only line 4, while variable \texttt{provider} is created in block $B$ from lines 2--5, and $create\_dis($\texttt{provider}$, r)=1$. 
In general, $scope\_dis(B_1,B_2)=null$ if $B_1$ does not (in)directly contain $B_2$. In Definition \ref{dec_r_def}, since $c$ is accessible for the given context, $B'$ must be contained by $B$, and $create\_dis(c,r) \neq null$.
For the candidates which are accessible fields, $B$ is considered as the block containing all the blocks in the containing method. The closest block $B$ where a global variable is created can be considered as the outermost block.

\textbf{Accessing-recentness}. The access variable $c$ could be used in the code blocks/regions which have no ordering or containing relation, e.g different methods. Thus, \textit{accessing-recentness} should be estimated within methods and not by using scope-distance. 
%
In \tool, the \textit{accessing-recentness} of candidate $c$ is simply calculated as the distance in lines of code (LOC) as follows:
\begin{definition}{\textbf{(accessing-recentness).}}
For an AR request $r=\langle P, m\_call, pos \rangle$ and a candidate $c$, the \textit{accessing-recentness} $access\_rec(r,c)$ is the distance in LOC between the line (indexed $L$) where $m\_call$ is called and the line (indexed $L'$) where $c$ is most recently used/defined in the containing method, $access\_rec(c,r) = |L-L'|$.
\end{definition}
For AR request $r$ in Figure \ref{fig:scope_code}, $access\_rec($\texttt{user}$, r)=0$ and $access\_rec($\texttt{provider}$,r)=1$. Note that if $c$ is never defined and used in the containing method, then $access\_rec(c,r) = null$. For the candidates which are fields or global variables, the methods, which define/use them, could be invoked in the containing method. An inter-procedural analysis can be performed to more accurately measure $access\_rec$. However, those methods can call other methods. Thus, a (recursive) inter-procedural analysis with a certain depth is required. For an efficient estimation of $access\_rec$, we set the depth as $0$ in this version of \tool.

\begin{figure}[h!]
\centering
\begin{subfigure}{.25\textwidth}
  \centering
  \includegraphics[width=1\linewidth]{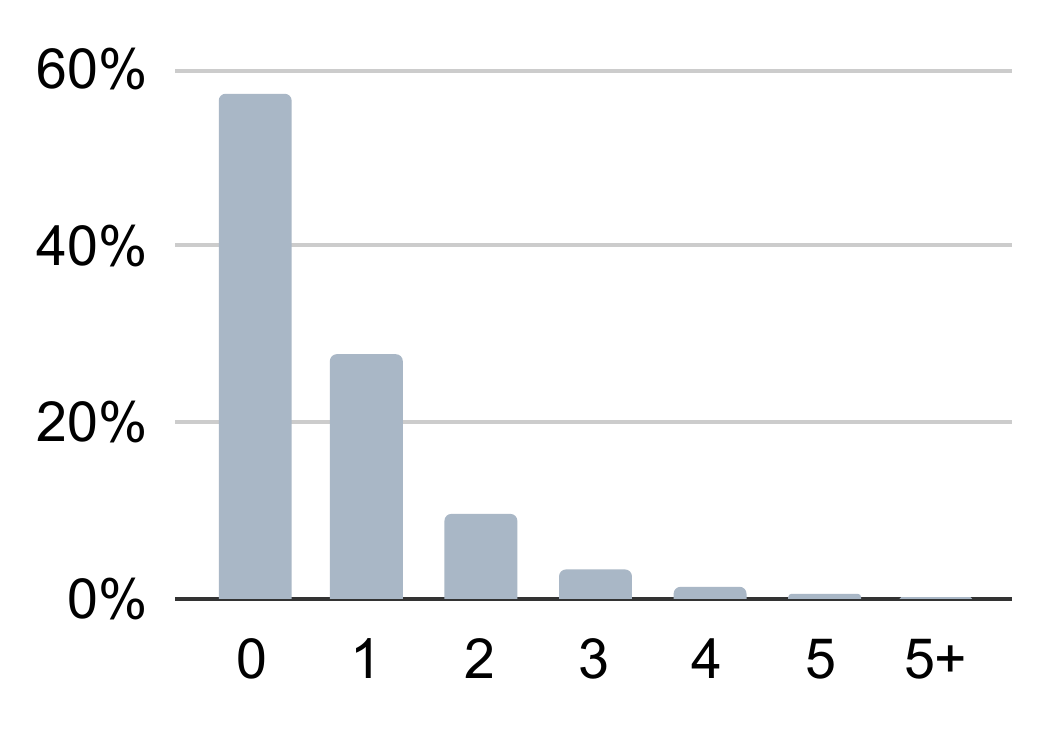}
  \caption{}
  \label{fig:dec_re}
\end{subfigure}%
%
%
\begin{subfigure}{.25\textwidth}
  \centering
  \includegraphics[width=1\linewidth]{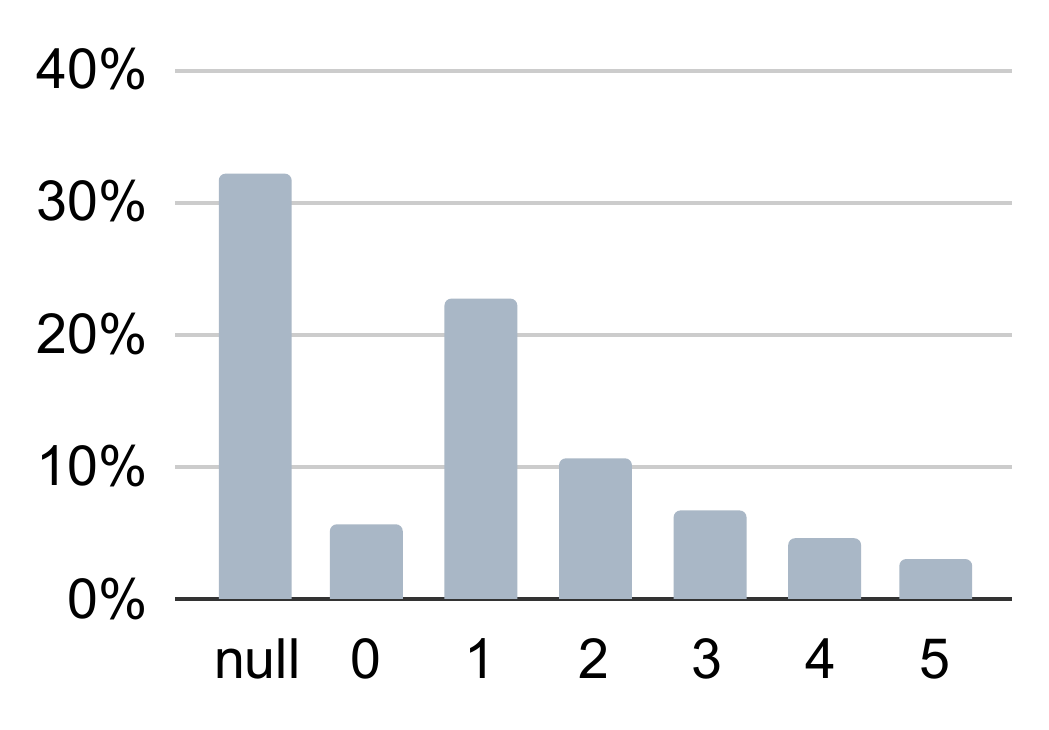}
  \caption{}
  \label{fig:use_re}
\end{subfigure}
\caption{(a) \textit{creating-distance}; (b) \textit{accessing-recentness}}
\label{fig:dec_use_re}
\end{figure}

In this version of \tool, the \textit{creating-distance} and \textit{accessing-recentness} of candidates are computed based on heuristics. In fact, they can be computed more accurately by applying data-flow analyses. However, as the program to be completed is often not compilable, partial analysis techniques such as PPA~\cite{ppa} could produce unreliable results, especially for complex code. Additionally, an accurate analysis might be costly. Meanwhile, the \textit{creating-distance} and \textit{accessing-recentness} of candidates, which are estimated by the current simple method, can still capture the recentness of the candidates which are variables. 
%
Indeed, Figure \ref{fig:dec_use_re} shows the distribution of the \textit{creating-distance/accessing-recentness} of the arguments which are a variable (\textit{variable arguments}) in our large corpus.
As seen, the variable arguments tend to be ``\textit{closer}'' to the requesting position. 
For \textit{creating-distance}, nearly 3/5 variable arguments are declared in the same scope as their corresponding method calls (\textit{creating-distance} is 0). 
Meanwhile, the variable arguments are often either used for the first time in the containing method (\textit{accessing-recentness} is $null$), or reused very quickly (within a few LOCs). However, reusing variables at the same line is less likely.
From these results, we conclude that for variable arguments, developers usually use the variables recently declared or defined/used. Thus, \textit{creating-distance} and \textit{accessing-recentness} should be used as recommending features in \tool's AR process. 
%



\begin{figure}[]
\centering
\begin{lstlisting}[language=Java]
public int insert(User user){
  String provider = user.getProvider();
  if(StringUtils.equals(``standard``, provider)){
        ignoreProvider(user,/* AR_REQUEST */);
  }
}
\end{lstlisting}

\caption{An example in \texttt{xdiamond} project}
\label{fig:scope_code}
\end{figure}

\section{\tool: An Effective API Argument Recommendation Approach}
\label{sec:argrec}

\begin{figure*}[h]
    \centering
    \includegraphics[width=\textwidth]{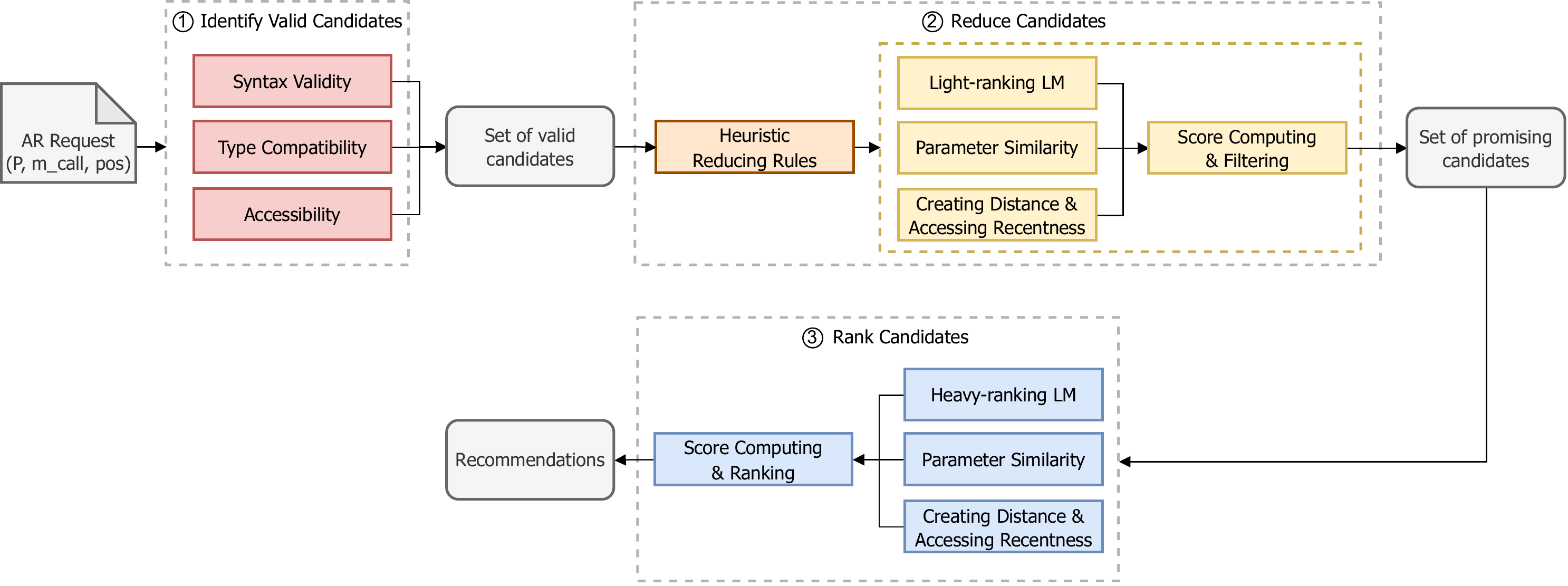}
    \caption{\tool's API Argument Recommendation Process Overview}
    \label{fig:overview}
\end{figure*}

Figure \ref{fig:overview} illustrates the argument recommendation (AR) process of \tool. For an AR request, \tool identifies the set of valid argument candidates which are syntax-valid, type-compatible, and accessible (\textit{valid}) for the given context. Next, the set of valid candidates is reduced by filtering out unpromising candidates. This could improve the candidate ranking performance of the final step in not only accuracy but also recommending time.


\subsection{Identifying Valid Candidates}
Given an AR request $r=\langle P, m\_call, pos \rangle$, \tool analyzes the partially completed code $P$ to generate a set of code expressions based on the syntax, type-compatibility, and accessibility constraints defined by the  programming language in use such as Java. The main idea is that from accessible expressions, \tool constructs more complex accessible expressions whose data types are compatible with the expected data type(s) $t$. The generation starts with the accessible variables/methods, constants, static methods, and classes in the given partially completed code $P$.

First, the expected data type $t$ of the $pos^{th}$ parameter in the method declaration corresponding to $m\_call$ is identified. In \tool, to avoid misleadingly omitting valid candidates, all the overloading method declarations are also considered. For example, both \texttt{Pos getPosition(Character)} and \texttt{Pos getPosition(String)}, which are the members of the same class in the context, are considered for the request $r=\langle P,~$\texttt{getPosition}$, 1 \rangle$. In this case, $t$ could be either \texttt{Character} or \texttt{String}.

Next, the initial set of accessible expressions is constructed by identifying all the accessible variables, methods, constants, static methods, and classes in the given context $P$. These expressions are used to generate the more complex expressions based on the grammar and accessibility constraints. 
Essentially, valid expressions could be infinitely complex.
However, most of the arguments belong to a small number of expression types in practice~\cite{Precise,java_spec}. As shown in Table \ref{tab:expr_types}, 99.43\%  of the arguments in our dataset in Section \ref{sec:example} are categorized into 17 types of expressions. Among these types, 15/17 types are supported by \tool, we leave out two very complex expression types  \textit{lambda expression} and \textit{compound expression}. These complex expressions are too costly to be generated and account for only about 4\% of the arguments. As a result, 95\% of all the arguments in our dataset are covered by \tool. 
Finally, the generated expressions are checked if their data type is compatible with at least one expected data type of $r$. Particularly, the data type $t'$ of a valid candidate must be \textit{compatible} with $t$ without casting. In other words, $t=t'$ or $t'$ inherits from $t$.

\begin{table*}[]
\small
\caption{Argument expression types and their grammar. This grammar just shows the types' structures, and the accurate grammar can be found in Java Specification~\cite{java_spec}.}
\label{tab:expr_types}
\begin{tabular}{@{}l|l|r@{}}
\toprule
Expr. type & Grammar & Distr. (\%) \\ \midrule
$\langle$simple name$\rangle$        & $\langle$identifier$\rangle$              & 49.31               \\
$\langle$method invoc$\rangle$  & [$\langle$expr$\rangle$"."]$\langle$identifier$\rangle$"("$\langle$arguments$\rangle$")"              & 13.01               \\
$\langle$field access$\rangle$       & $\langle$expr$\rangle$"."$\langle$identifier$\rangle$              & 11.07               \\
$\langle$string lit$\rangle$         &--               & 4.91                \\
$\langle$number lit$\rangle$         &--               & 3.27                \\
$\langle$bool lit$\rangle$           &--               & 3.69                \\
$\langle$null lit$\rangle$           &--               & 1.99                \\
$\langle$char lit$\rangle$           &--               & 0.88                \\
$\langle$type lit$\rangle$           &--               & 0.71                \\
$\langle$lit$\rangle$                & $\langle$string lit$\rangle$|$\langle$bool lit$\rangle$|$\langle$null lit$\rangle$ & --       \\
                                     & |$\langle$number lit$\rangle$|$\langle$char lit$\rangle$|$\langle$type lit$\rangle$ & \\
$\langle$cast expr$\rangle$          & "("$\langle$type$\rangle$")"$\langle$expr$\rangle$              & 0.95                \\
$\langle$obj creation$\rangle$       & "new "$\langle$class type$\rangle$ "("[$\langle$arguments$\rangle$]")"              & 2.77                \\
$\langle$array creation$\rangle$     & "new "$\langle$type$\rangle$"["$\langle$dim$\rangle$"]"              & 0.71                \\
$\langle$compound expr$\rangle$      & $\langle$unary expr$\rangle$|$\langle$binary expr$\rangle$              & 3.68                \\
$\langle$this expr$\rangle$               &--               & 0.93                \\
$\langle$lambda expr$\rangle$        & ($\langle$lambda para list$\rangle$) $\langle$lambda body$\rangle$             & 0.75                \\
$\langle$array access$\rangle$       & $\langle$expr$\rangle$"["$\langle$index$\rangle$"]"              & 0.71                \\
$\langle$static method ref$\rangle$  & $\langle$class type$\rangle$::$\langle$identifier$\rangle$              & 0.09                \\
$\langle$expr$\rangle$               & $\langle$simple name$\rangle$|$\langle$method invoc$\rangle$|$\langle$cast expr$\rangle$|   & --  \\ 
                   & |$\langle$field access$\rangle$|$\langle$lit$\rangle$|$\langle$array access$\rangle$ & \\
                   \midrule

Total              &               & 99.43               \\ \bottomrule
\end{tabular}
\end{table*}

Without loss of generality, the \textit{arguments} of \textit{method invocation} and \textit{object creation}, the \textit{dim} of \textit{array creation}, and the \textit{index} of \textit{array access} can be treated as new arguments. Thus, these elements are left for the next AR requests of users in \tool to avoid infinite argument generation. 
For the example shown in Figure~\ref{fig:example_filtering}, the request expects an \texttt{Object}, and after apply our candidate identification method, there are about 40K valid candidates. To improve AR performance, it's essential to narrow down such very large sets of valid candidates. In the next section, we introduce our method to reduce the identified candidate sets.

\subsection{Reducing Candidates}

\label{sec:filter_cand}
For a request $r=\langle P, m\_call, pos \rangle$, the set of valid candidates could be very large, especially when $r$ expects an argument of a common data type such as \texttt{Object} or \texttt{String}. 
%
%
%
Meanwhile, as only one candidate among them must be recommended efficiently to complete the method call, the very large candidate set could cause incorrect recommendations. Moreover, it could become too expensive for heavy-ranking models (e.g., DNN models) to evaluate all the valid argument candidates. The reason is that these LMs could be generative models which are not designed to rank a large set of candidates.
In \tool, we develop a set of heuristic rules and apply a light-ranking step to efficiently reduce the valid candidate set by filtering out unpromising candidates before passing them to the heavy-ranking stage. 

\subsubsection{Candidate Reducing by Heuristic Rules}
\label{staticparamfilter}
In large projects such as Eclipse or Netbeans, their sets of static methods or static variables (public static fields) are usually very large. For example, Eclipse has about +74K static methods/variables, while this figure for Netbeans is +52K. Since static methods/variables and static methods can be used everywhere (i.e., always accessible), these large sets of static methods/variables could cause huge sets of valid candidates (e.g., Figure \ref{fig:example_filtering}). 
%
Thus, we design the heuristic rules in \tool to reduce the subset of candidates which are generated from static methods/variables. 
The valid candidate $c$ generated from static methods/variables \textbf{must satisfy at least one} (i.e., \textit{OR} relation) of the reducing rules to be kept in the reduced candidate set.

\textbf{Rule of shared sub-tokens}. 
Variable $c$ shares at least a subtoken (case-insensitive) with the names of either the calling method, the name or data type of the calling object, or the containing method.
For a method $m$ called by object $o$ in the containing method $M$, the arguments of $m$ usually contain certain subtokens in the names of $m$, $o$, and $M$, and the data type of $o$. 
In Figure~\ref{fig:example_filtering}, both the expected argument's name, \texttt{PROP\_BACKGROUND\_DRAWING} and the called method's name \texttt{getProp} also contains sub-token \texttt{prop}. 
Both the containing method's name, \texttt{isBackgroundDrawing}, and the argument's name also contain \texttt{background}.
%
%
In our dataset, we found that 31.3\% of the arguments share at least a subtoken with the names of either the called method, the name or data type of calling object, or the containing method.
For Figure \ref{fig:example_filtering}, 
this rule effectively and correctly reduces the candidate set to only 3.5K candidates from more than 52K valid candidates.

\textbf{Rule of recently-used classes}. 
Variable $c$ is a member of a class which is recently used to access its members (fields/methods) in the coding context. The rule ensures that \tool never misses the static methods/variables or static methods in the class it has been encountered. 

Besides these rules, we also apply other rules, such as:
\begin{itemize}
    \item $c$ is a boundary variable (e.g., max/min).
    \item $c$ is a member of the class/parent classes of the context.
    \item $c$ is declared in the same package scope of the context.
\end{itemize}
We have preliminarily evaluated those rules on +35K AR requests in 10\% randomly selected files in Netbeans. The results show that 91.5\% of (unpromising) candidates have been eliminated without sacrificing much accuracy, i.e., there are only about 2\% of the requests where the expected arguments are incorrectly eliminated.

\subsubsection{Candidate Reducing by Light-ranking Model}
Although the reducing rules filter out a large portion of valid yet unpromising candidates, the number of remaining candidates might still be large. For AR, a time-sensitive task, this could create difficulties in applying powerful (yet computationally expensive) models to quickly evaluate and rank these candidates.  
To further narrow down the search space, we select top-$RT$ candidates ranked based on their \textit{parameter similarity}, \textit{creating-distance}, \textit{accessing-recentness} and the evaluation of a light-ranking language model. Specifically, a light-ranking model should be computationally efficient language model. In this work, we use a count-based method such as $n$-gram LM~\cite{n_gram} or nested-cached LM~\cite{nestedcache} for this light-ranking stage.
%

%
$RT$ is the predefined \textit{reducing threshold} and should be able to balance between candidate reduction performance and accuracy sacrifice (incorrectly eliminating expected arguments). 
%
Specially, given an AR request $r=\langle P, m\_call, pos \rangle$, the ranking score of a candidate $c$ is computed as:
\begin{equation}
\label{score_lr}
score_{lr}(c,r) = \sqrt[1+v]{\mathcal{P}_{lr}(c,P) \times parasim(c,p) \times  recent(c,r)^v}
\end{equation}
where $p$ is the parameter's name, and $\mathcal{P}_{lr}(c,P)$ is the likelihood that $c$ is the next sequence following $P$, which is evaluated by a lightweight LM trained on the lexical form of a corpus. In formula (\ref{score_lr}), $parasim(c, p)$ is the \textit{parameter similarity} of $c$. Since the features of recentness are only applied for variables, if $c$ is not a variable then $v=0$, and only $\mathcal{P}_{lr}$ and $parasim$ are considered in $score_{lr}$. Otherwise, $v=1$ and $recent(c,r) = Prob(D=create\_dis(c,r)) \times Prob(U=access\_rec(c,r))$, such that $Prob(D=d)$ and $Prob(U=u)$ are the probability in a dataset that \textit{creating-distance} equals to $d$ and \textit{accessing-recentness} equals to $u$. Setting $v=1$ when $c$ is variable and $v=0$ otherwise also makes $score_{lr}$ comparable in these two cases.
%
Candidate $c$ is ranked higher if $c$ is more similar to the recommending parameter's name, ``closer'' to the requesting position, and more likely to be the next sequence following $P$.

\textit{In summary}, the set of the valid candidates is reduced by the heuristic reducing rules. The reduced candidate set is further narrowed down efficiently up to a reducing threshold, $RT$, which remains the  most promising candidates. To isolate top-$RT$ promising ones, the candidates are evaluated based on their similarity with the recommending parameter, recentness, and likelihood to be the next sequence following the coding context.

\subsection{Ranking Candidates}

To produce recommendations, \tool ranks the reduced candidate set based on their probability to be the argument of the calling method. In \tool, the probability of a candidate is evaluated by a heavy-ranking model combined with the parameter similarity and recentness of the candidate.
Compared to the count-based models (e.g., $n$-gram LM) in the light-ranking stage, the heavy-ranking models applied in this stage should be capable of leveraging longer dependencies. The suitable options for heavy-ranking models could be neural network LMs such as LSTM, GPT-2, or CodeT5.

\textbf{Selective Prediction}. 
In fact, an LM could be more effective for arguments in certain expression types. To verify this assumption, we have performed an experiment to preliminarily evaluate the performance of the trained $n$-gram LM with nested-cached~\cite{nestedcache} and GPT-2~\cite{gpt2} in recommending arguments of method calls of Lucene project\footnote{https://github.com/apache/lucene}. The results show that GPT-2 is better than $n$-gram LM with nested-cached in the cases whose the argument expression types are \textit{Simple Name}, \textit{Array Access}, \textit{Type Literal}, \textit{Object Creation}, \textit{Array Creation}, \textit{Lambda Expr}, and \textit{Method Reference}. For the cases with the remaining expression types, $n$-gram with nested-cached performed better.
%
%
%
Thus, given an AR request $r=\langle P, m\_call, pos \rangle$, we adopt ensemble methods~\cite{ensemble} in \tool to compute the likelihood of a candidate $c$ to be the next sequence following $P$:
\begin{equation}
\label{selective_pred}
\mathcal{P}(c,P)=
\begin{cases}
\mathcal{P}_{hr}(c,P) & expr\_type(c) \in E \\
\mathcal{P}_{lr}(c,P) & \text{Otherwise}
\end{cases}
\end{equation}
where $\mathcal{P}_{lr}(c,P)$ and $\mathcal{P}_{hr}(c,P)$ are evaluated by a lightweight LM and a heavy-weight LM, respectively. In formula (\ref{selective_pred}), $E$ is the set of expression types where the heavy-ranking model performs better than the light-ranking model. $E$ is empirically computed beforehand by evaluating the performance of the models, or $E$ can be gradually updated while being applied. As only the candidates of certain expression types are evaluated by the heavy-ranking model, this method reduces the workload for the heavy-ranking model in \tool while potentially improving the overall performance. 

\textbf{Combining score}. 
Given a request $r=\langle P, m\_call, pos \rangle$, the final ranking score of a candidate $c$ to be the argument for $r$ is computed as:
\begin{equation}
\label{overall_score}  
score(c,r) = \sqrt[1+v]{\mathcal{P}(c,P) \times parasim(c,p) \times  recent(c,r)^v}
\end{equation}
where $p$ is the parameter's name. Similar to $score_{lr}$, in formula (\ref{overall_score}), if $c$ is a variable, $v=1$ and $recent(c,r) = Prob(D=create\_dis(c,r)) \times Prob(U=access\_rec(c,r))$, such that $Prob(D=d)$ and $Prob(U=u)$ are the probability in a dataset that \textit{creating-distance} equals to $d$ and \textit{accessing-recentness} equals to $u$. 
If $c$ is not a variable, $v=0$.
Finally, the candidates are ranked  according to their $score(c,r)$ score.

\section{Empirical Methodology}
\label{sec:empirical_methodology}
We seek to answer the following research questions:
\begin{itemize}[leftmargin=*]
    \item \textbf{RQ5: \textit{Accuracy Comparison}.} How accurate is \tool in argument recommendation (AR)? How is it compared with the state-of-the-art AR approaches~\cite{PARC,nestedcache,tabnine}?
    \item  \textbf{RQ6: \textit{Sensitivity Analysis}.} How do various factors affect \tool, e.g. argument recommendation scenarios, training data's sizes, context lengths?
    \item  \textbf{RQ7: \textit{Intrinsic Analysis}.} How do \tool's components, e.g., candidate identification, candidate reduction, heavy-ranking model, and reducing threshold, contribute to its performance?
    \item  \textbf{RQ8: \textit{Time Complexity}.} What is our running time?
\end{itemize}




\subsection{Datasets}                    

\begin{table}[]
\small
\centering
\caption{Statistics of the datasets}
\label{tab:datasets}
\begin{tabular}{@{}l|r|r@{}}
\toprule
                    & Small corpus & Large corpus \\ \midrule
\#Projects          & Eclipse \& Netbeans   & 9,271       \\
\#Files             & 53,787       & 961,493      \\
\#LOCs              & 7,218,637    & 84,236,829   \\ 
\#AR requests       & 700,696    & 913,175   \\ \bottomrule
\end{tabular}
\end{table}
In this study, we used two datasets for evaluation, \textit{small corpus} and \textit{large corpus} (Table \ref{tab:datasets}). For comparison, we used the same dataset (\textit{small corpus}) as in the state-of-the-art technique for AR~\cite{PARC}. The \textit{small corpus} contains two large projects: Eclipse and Netbeans. This corpus contains a relatively large code base, about 54K files and 7.2M LOCs. 
For further evaluation, we used the \textit{large corpus} by Allamanis \textit{et al.}~\cite{allamanis_data} containing  +9.2K popular Java projects from Github. The large corpus is widely used to evaluate the existing code completion approaches~\cite{bigcode,curglm,nestedcache}. 
For a non-bias evaluation, these datasets in this evaluation are different from the dataset in our empirical study (Section \ref{sec:example}).


%

\subsection{Evaluation Setup, Procedure, and Metrics}

\textbf{RQ5. Accuracy Comparison}

\textit{\underline{Baselines}.} We compare \tool with the state-of-the-art argument recommendation (AR) techniques: 
(1) \textbf{PARC}~\cite{PARC}, an IR-based AR approach and specialized for AR, represents argument usages by several static features, constructs a usage database, and searches for similar usages when requested; 
(2) \textbf{GPT-2}, the core engine of Tabnine~\cite{tabnine} and IntelliCode~\cite{intellicode}, two of the most popular AI-assisted code completer;
(3) \textbf{CodeT5}~\cite{codet5}, a encoder-decoder model specialized for source code;
(4) \textbf{SLP}~\cite{nestedcache}, a representative code completer is carefully adapted from $n$-gram models for source code and shown that it can even better than RNN and LSTM based models~\cite{nestedcache}.
For each baseline, we applied a post-processing step to exclude the candidates which are type-incompatible in the recommendation sets resulted by the code model. 
Note that there are several others powerful models designed for general SE tasks such as CodeBERT~\cite{codebert} or CuBERT~\cite{cubert} which could be applied for the AR task. However, for a practical evaluation, we select CodeT5~\cite{codet5}, which is one of the most recent and powerful models specialized for code, as a baseline.
%


\textit{\underline{Procedure}.} Since the source of PARC is not publicly available, for a fair comparison with PARC, we use the original results of PARC~\cite{PARC}, and apply the same dataset (the \textit{small corpus}) and evaluation setting for the other approaches. Specially, we randomly shuffle and divide the source files of each project into 10 equal folds. We perform 10-fold cross-validation: each fold was chosen for testing, and the remaining folds were used for training. Since PARC focuses on AR for \textit{frequently-used API libraries}~\cite{PARC}, it fits well with recommending certain API methods in the common libraries such as SWT, AWT, or Java Swing which are frequently used in developing applications.

Meanwhile, \gpt, CodeT5, and SLP can be applied to recommend arguments for a wider range of APIs/method calls. However, \gpt, CodeT5, and SLP are designed to predict next tokens for the given context. In this work, we adapt \gpt, CodeT5, and SLP for AR and evaluate the AR performance of \tool and the two baselines in the \textit{general AR task}, i.e., recommending arguments for any method call. 
%
%
To adapt these approaches for AR task, we apply the same mechanism used by Karampatsis \etal~\cite{bigcode} to combine multiple predicted tokens to form a whole argument. Particularly, while predicting next tokens, the beam search algorithm is applied to find the \textit{k} highest probability sequences of tokens, which each of them corresponds to a recommended argument.

\textit{\underline{Metrics}.} To compare with PARC~\cite{PARC} in recommending arguments for method in \textit{frequently-used libraries}, we adopt the same metrics used in PARC, \textit{Precision} and \textit{Recall} at top $k$. Specially, $Precision(k) =\frac{R(k)}{S}$ and $Recall(k) =\frac{R(k)}{A}$, where $A$ is the total number of AR requests in the test set, \textit{S} is the number of requests where the expression type of the expected argument is supported by the tools, and $R(k)$ is the number of requests where the recommended argument in top-$k$ of the ranked list matches the expected argument.
%
%
%
%

Note that an expected argument could be a literal or a lambda expression. AR tools might not be able to correctly recommend the whole expressions~\cite{ase19}. Thus, if the approaches can correctly suggest the expression type, we consider it as a correct recommendation. Instead of suggesting incorrect ones, recommending default values (e.g., \texttt{"<EMPTY\_STRING>"} for string literals and or \texttt{0} for number literals) or a template of the expected expression types would be more appropriate for these cases. This is still useful for developers and can be done by a post-processing step.

For \textit{general AR task} in the other experiments, we apply top-$k$ accuracy (aka., \textit{Recall} at top-$k$), and \textit{Mean Reciprocal Rank (MRR)}, which are widely used in code completion~\cite{bigcode, nestedcache}. Top-$k$ accuracy represents whether the target arguments are presented in top-$k$ ranked lists or not. \textit{MRR} is the average of reciprocal ranks for the test set,
$
    MRR = \frac{1}{A} \sum_{i=1}^{A}{\frac{1}{rank_i}}
$,
where $A$ is the total number of AR requests in test set, and $rank_i$ is the position of expected argument in the ranked list of request $i^{th}$ in the test set.
\textit{MRR} focuses more on the rank of target argument in each suggestion list. For example, an \textit{MRR} value of 0.5 shows that AR technique could suggest correct arguments at position 2 of ranked lists on average.

\textit{\underline{Evaluation Setup}.} 
\textit{For \tool}, $n$-gram LM with nested-cache is used for the light-ranking model, the reducing threshold $RT = 20$, GPT-2 is used for heavy-ranking model, the sets of expression types selective prediction and recentness probabilities are derived by analyzing the dataset used in Section \ref{sec:example} which can be found in our website~\cite{ourwebsite}. $n$-gram LM uses the Jelinek-Mercer smoothing with a fixed confidence of 0.5 and $n=6$. For detailed implementation, see our website~\cite{ourwebsite}.

\textit{For SLP}, we use their default settings in training and predicting for the evaluation\footnote{https://github.com/SLP-team/SLP-Core}. 
\textit{For GPT-2}, we use the pre-trained model with 345M parameters by OpenAI\footnote{https://openai.com/blog/tags/gpt-2}. 
%
%
\textit{For CodeT5}, we use the pre-trained \textit{CodeT5-base} model with 220M parameters released by Salesforce\footnote{https://blog.salesforceairesearch.com/codet5/}. In our experiments, all the pre-trained models are fine-tuned on the training data.

In this paper, all our experiments were run on a server with Intel(R) Xeon(R) CPU @ 2.30GHz, 32GB RAM, and Tesla P100 GPU.

\noindent\textbf{RQ6. Sensitivity Analysis}.


In \tool, the heavy-ranking model performs complex computations and should be deployed on a powerful server as a central/public resource for all users. Thus, this model should \textit{arguably} not learn from a user's local source files and use the learned knowledge for others. 
Meanwhile, the light-ranking model which is very light-weight can be deployed in developers' (much less powerful) computer. 
This enables \tool to \textit{personalize} the argument recommendations by learning developers' coding practice from their local source files. As a result, \tool can better support developers in several scenarios. In this work, we aim to investigate the impact of 3 different real-world AR scenarios on \tool's performance on the large corpus:

\begin{itemize}[leftmargin=*]
    \item \textit{New-project}. This setting reflects the scenario when \tool encounters completely new projects in practice. The tool learns from a fixed training set and then is tested on the separate testing set for evaluating performance. This usage is cross-project: 400 projects (about 5\% of the projects) for testing and the remaining for training. 
    \item \textit{Working-project}. This scenario comes from the idea that when developers need argument recommendations from the tool for the projects that they are working on. In this scenario, some source files available at that time are used for \textit{personalizing} the argument recommendation task. Specially, instead of fixing the training set like \textit{new-project} setting, \tool could iteratively learn after each recommendation during testing. In this setting, the light-ranking model, which could be deployed on developers' machine, can learn the their coding practice from the available source files.
    \item \textit{Maintaining-project}. This setting simulates the scenario that developers make AR requests when they maintain their projects. In this scenario, developers usually make small changes (e.g., several files/methods) to the existing code. The tool is tested on one file at a time in the test set. \tool could learn from all the remaining files in the corpus. 
\end{itemize}
%
%
%
%
Besides, we also study the impacts of the following factors on the performance of \tool: \textit{training data size}, \textit{context length}, and \textit{origin of called methods}. 

\noindent\textbf{RQ7. Intrinsic Analysis}.

We study the impacts of the following components: valid candidate identification, candidate reduction, candidate ranking, reducing size, and static features. We create different variants of \tool with different combinations and measured their performance. Particularly, to evaluate the impact of a component, we create a variant where the component is disabled and the others are on.

\section{Empirical Results}
\label{sec:empirical_results}

\subsection{Accuracy Comparison (RQ5)}
\label{sec:accuray_comparison}
Table \ref{tab:small_corpus_parc} shows the comparison results of \tool against the state-of-the-art techniques in argument recommendation (AR), PARC~\cite{PARC}. For the results in Table \ref{tab:small_corpus_parc}, we applied the same evaluation procedure used in PARC~\cite{PARC}. Particularly, the techniques are tested on recommending arguments for method calls of \textit{frequently-used libraries}: SWT for Eclipse and Java Swing and AWT for Netbeans. 
As seen, \textit{\tool outperforms PARC in both \textit{Precision} and \textit{Recall} for both Eclipse and Netbeans}.  
For Netbeans, \tool achieves 16\% and 15\% better in the \textit{Precision} and \textit{Recall} than PARC at the top position. With Eclipse, both the \textit{Precision} and \textit{Recall} at the top position of \tool are also much better than PARC, 19\% and 18\%. With higher \textit{Recall} and \textit{Precision}, \tool is able to recommend more diverse argument expression types as well as recommends arguments more precisely.

Analyzing the results, we found the main reason for the higher accuracy of \tool compared to PARC. 
%
%
Particularly, arguments are quite unique in practice (Section \ref{sec:example}), and \tool analyzes the code context to identify accessible and type-compatible argument candidates and learns to recommend arguments. Meanwhile, PARC retrieves the candidates stored in a database. Thus, \tool can give the correct recommendations for the requests, even when their expected arguments have not been seen, while PARC cannot.

\begin{table}[h!]
\small
\centering
\caption{Performance of \tool and PARC in AR task for the methods in the \textit{frequently-used libraries}}
\label{tab:small_corpus_parc}
\resizebox{\columnwidth}{!}{%
\begin{tabular}{llcccc}
\toprule
\multirow{2}{*}{Project} & \multirow{2}{*}{} & \multicolumn{2}{c}{\tool} & \multicolumn{2}{c}{PARC} \\
\cmidrule(lr){3-4} \cmidrule(l){5-6}
 &  & \textit{Precision} & \textit{Recall} & \textit{Precision} & \textit{Recall} \\
 \midrule
\multirow{3}{*}{Netbeans} & Top-1 & 52.92\% & 51.67\% & 46.46\% & 44.86\% \\
 & Top-3 & 70.18\% & 68.28\% & 66.20\% & 66.75\% \\
 & Top-10 & 78.36\% & 76.15\% & 72.06\% & 69.57\% \\
 \midrule
\multirow{3}{*}{Eclipse} & Top-1 & 56.66\% & 55.04\% & 47.65\% & 46.65\% \\
 & Top-3 & 67.88\% & 65.63\% & 65.05\% & 63.68\% \\
 & Top-10 & 73.14\% & 70.76\% & 72.26\% & 70.73\% \\
\bottomrule
\end{tabular}%
}
\end{table}

For \textit{general AR task} (i.e., recommending argument for every method call), Table \ref{tab:small_corpus_top_k} shows the performances of \tool, \gpt, CodeT5~\cite{codet5}, and SLP~\cite{nestedcache}. As seen, \textit{\tool's performance is significantly better than those of \gpt, CodeT5, and SLP}. Particularly, compared to \gpt, the top-1 accuracy of \tool is 24\% and 14\% better for Netbeans and Eclipse. The top-3 accuracy of \tool is 16.4\% and 13.5\% better for those of CodeT5. For SLP, the improvements are much more significant, 87\% and 125\% for Netbeans and Eclipse. 
Especially, the top-5 accuracy of \tool is about 80\% on average. This means for 4/5 requests, the expected arguments can be found at top-5 positions of the recommended lists of \tool.
Plus, \textit{MRR} of \tool is about 0.71. In other words, \tool correctly recommended the expected arguments from the first to the second positions on average. This is about 12\%, 24\%, and 73\% relatively better than \gpt, CodeT5, and SLP.

\begin{figure}
  \centering
\begin{lstlisting}[language=Java]
// Package: org.netbeans.editor.ext
public class FormatSupport {
    public TokenItem insertToken(TokenItem beforeToken, TokenID tokenID, TokenContextPath tokenContextPath, String tokenImage) {...}
}
// Package: org.netbeans.editor
public interface ImageTokenID extends TokenID {
    public String getImage();
}
// Package: org.netbeans.editor.ext
public class ExtFormatSupport extends FormatSupport {
    public TokenItem insertImageToken(TokenItem beforeToken, ImageTokenID tokenID, TokenContextPath tokenContextPath) {
        return super.insertToken(beforeToken, tokenID, tokenContextPath, /* AR_REQUEST */);
                    //Expected a String: tokenID.getImage()
    }
}
\end{lstlisting}
  \caption{An argument recommendation request in Netbeans}
  \label{fig:result_example}
\end{figure}%

Analyzing the cases where \tool performed better than \gpt, CodeT5, and SLP, we found that even having high accuracy in predicting next tokens, these techniques failed to construct the expected arguments which contain multiple tokens. This causes their lower accuracy compared to \tool's.
For example, for the AR request at line 12 in Figure \ref{fig:result_example}, GPT-2, CodeT5, and SLP can produce the recommendations which are partially correct. The closest recommendation of CodeT5 and GPT-2 is \texttt{tokenContextPath}. However, this recommendation is type-incompatible 
%
(\texttt{TokenContextPath} vs \texttt{String}). 
Thus, this candidate is filtered out from the candidate list, and the remaining ones such as \texttt{text} are even less relevant.
SLP suggested variable \texttt{tokenImage} which is unavailable in the context, i.e., the variable is not declared in \texttt{insertImageToken}, and neither \texttt{ExtFormatSupport} nor \texttt{FormatSupport} has any field named \texttt{tokenImage}.
Meanwhile, \tool generated the valid candidate set including \texttt{tokenID.getImage()} in the context before feeding them to the LMs. Because of the likelihood and the \textit{parasim} of \texttt{tokenID.getImage()}, the candidate is ranked first by \tool.


\begin{figure}
  \centering
\begin{lstlisting}[language=Java]
public final class DebuggerEngine implements ContextProvider {
    public class Destructor {
        //...
    }
}
public final class DebuggerManager implements ContextProvider {
    public DebuggerEngine[] startDebugging (DebuggerInfo info) {
        //...
        ep.setDestructor (/* AR_REQUEST */);
                    //Expected: engine.new Destructor ()
    }
    
}
\end{lstlisting}
  \caption{An example which argument recommendation tools often fail to suggest correct argument}
  \label{fig:failed_case}
\end{figure}%

By analyzing the cases where \tool did not work well (i.e., cannot rank the expected arguments at the top-3 positions), we found that one of the main reasons is, that the expected expression types are unpopular and not supported by \tool. Consequently, \tool cannot construct and recommend the correct arguments.
Figure~\ref{fig:failed_case} shows an example in which \tool failed to suggest the correct argument. In this case, \texttt{setDestructor} requires an object type \texttt{Destructor} (line 9).
This is a challenging case, because to fill a correct argument, a new object type \texttt{Destructor} should be created and passed to the method. 
Meanwhile, \texttt{Destructor} is a non-static inner class of \texttt{DebuggerEngine}, and to initialize a \texttt{Destructor} object, an object of the outer class (\texttt{engine}) must be used. This argument expression type is very unpopular and not supported by \tool. Consequently, \tool fails to generate a correct candidate for this challenging case. Also, for this complicated case, \gpt, CodeT5, and SLP cannot suggest the expected argument.

\begin{table}[!h]
\centering
   
\caption{Comparison in \textit{general AR task}}
\label{tab:small_corpus_top_k}
\resizebox{\columnwidth}{!}{%
\begin{tabular}{@{}llcccc@{}}
\toprule
Project & & \multicolumn{1}{c}{\tool} & \multicolumn{1}{c}{\gpt} & \multicolumn{1}{c}{CodeT5} & \multicolumn{1}{c}{SLP} \\
\midrule
\multirow{5}{*}{Netbeans} & Top-1 & 65.15\% & 52.63\% & 59.97\% & 34.91\% \\
 & Top-3 & 78.16\% & 57.69\% & 67.16\% & 48.10\% \\
 & Top-5 & 81.10\% & 57.87\% & 67.57\% & 55.02\% \\
 & Top-10 & 83.53\% & 57.88\% & 67.60\% & 67.20\% \\\cmidrule(l){2-6}
 & \textit{MRR} & 0.72 & 0.55 & 0.63 & 0.44 \\
 \midrule
\multirow{5}{*}{Eclipse} & Top-1 & 64.19\% & 56.53\% & 61.20\% & 28.52\% \\
 & Top-3 & 76.29\% & 61.89\% & 67.21\% & 41.60\% \\
 & Top-5 & 79.23\% & 62.09\% & 67.53\% & 49.46\% \\
 & Top-10 & 81.65\% & 62.10\% & 67.54\% & 62.67\% \\\cmidrule(l){2-6}
 & \textit{MRR} & 0.70 & 0.59 & 0.64 & 0.38\\
 \bottomrule
        \end{tabular}%
  }
\end{table}

\begin{table}[!h]
 \small
\centering
\caption{Top-$k$ accuracy of \tool in different scenarios}
\label{tab:impact_usages}
\begin{tabular}{lccc}
\toprule
 & New & Working & Maintain. \\
 & project & project & project \\
 \midrule
Top-1 & 53.42\% & 69.96\% & 74.49\% \\
Top-3 & 61.50\% & 81.14\% & 83.23\% \\
Top-5 & 64.21\% & 83.74\% & 85.38\% \\
Top-10 & 67.96\% & 85.88\% & 87.38\% \\ \midrule
\textit{MRR}     & 0.58   & 0.76  & 0.79  \\
\bottomrule
\end{tabular}%
\end{table}
\subsection{Sensitivity Results (RQ6)}
\label{sec:sensitivity}

\subsubsection{Impact of argument recommendation scenarios}

Table \ref{tab:impact_usages} shows the performance of \tool in different AR scenarios with the \textit{large corpus}. 
%
For newly-encountered projects (\textit{new-project} scenario),
\tool achieved an impressive performance. By \tool, developers can find their expected arguments at top-3 positions in 6/10 requests. 
Meanwhile, with \gpt which is the engine of Tabnine~\cite{tabnine}, the top-10 accuracy is only 56.1\%. 
%
For \textit{working} projects (\textit{working-project} scenario), with \textit{personalizing only the light-ranking model which can be deployed on developers' machines}, \tool achieved nearly 70\% top-1 accuracy and 0.76 \textit{MRR}. Especially, for \textit{maintaining} projects (\textit{maintaining-project} scenario), \tool can correctly recommend arguments at the top position in +7/10 requests.
\tool with personalization can adjust the parameters in its light-ranking model for the working projects to capture developers' coding practice and reduce the cases where the expected arguments are misleadingly eliminated during candidate reduction.
%
With this high accuracy, \textit{\tool can effectively assist developers in completing method calls while preserving their privacy.} 

\textbf{Unseen expected arguments}. We also studied the performance of \tool in recommending the expected arguments which have not been seen in the training corpus (\textit{unseen expected arguments}) for the \textit{new-project} scenario. 
%
We found that among the AR requests whose expected arguments is unseen, about 70\% of them are correctly suggested by \tool at the top-1 positions. 
%
This confirms our strategy of combining PA and LMs in the argument recommendation task. Specifically, PA is first used to identify correct arguments included in the valid candidate sets even when they are unseen, and then the LMs and the features specialized for AR are applied to learn recommending arguments rather than retrieving what has been stored in the training corpus.

\textbf{Accuracy by argument expression types}. Table  \ref{tab:top_1_expr_type} shows the performance of \tool by argument expression types. 
%
%
\begin{table}[]
\small
\centering
\caption{\tool's performance by the expected expression types}
\label{tab:top_1_expr_type}
\begin{tabular}{lrr}
\toprule
Expression type & \multicolumn{1}{l}{Distribution (\%)} & Top-1 (\%)\\
\midrule
Simple Name & 48.14 & 83.66  \\
Method Invocation & 15.19 & 45.51  \\
Field Access & 6.09  & 31.01  \\
Array Access & 0.74  & 53.26  \\
Cast Expr & 0.99  & 18.46  \\
String Literal & 10.03  & 98.14  \\
Number Literal & 5.06  & 95.66  \\
Character Literal & 0.47  & 87.93  \\
Type Literal & 0.90  & 81.92  \\
Bool Literal & 1.50  & 78.43  \\
Null Literal & 0.79  & 84.45  \\
Object Creation & 2.09  & 51.96  \\
Array Creation & 0.29  & 43.14  \\
This Expr & 1.06  & 91.05  \\
Super Expr & 0.00  & 0.00  \\
Compound Expr & 5.65  & 3.69  \\
Lambda Expr & 0.73  & 78.83  \\
Method Reference & 0.28  & 0.56  \\
\midrule
Total & 100.00  & 69.96  \\
\bottomrule
\end{tabular}%
\end{table}
As seen, \tool is very effective for arguments which are \textit{Simple Name} (e.g., variables or parameters), literals, or \textit{This Expr} expressions. Especially, for \textit{Simple Name} which accounts for a haft of the expected arguments, \tool can correctly rank the expected arguments at the top position in more than 8/10 requests. 
The reason is that the major portion of expected \textit{Simple Name} arguments are local variables or parameters of the containing methods. 
These arguments are correctly recommended because features \textit{parameter similarity}, \textit{creating-distance}, and \textit{accessing-recentness} are encoded in \tool. Meanwhile, \tool does not work very well for \textit{Method Invocation} and \textit{Field Access}, only 45.5\% and 31.0\%, respectively. 
However, this is still much better than the performance of \gpt, with about 29\% and 28\% top-1 accuracy for the same expression types. We found that among these AR requests, \gpt suggested invalid candidates in many cases. Also, \gpt prefers the candidates which are close to the requesting positions, even when they are inaccessible. Meanwhile, the expected arguments which are valid field access/method calls might not be very close to the positions or even defined in different files. 



\textbf{Accuracy by argument variable types}. We also investigated the performance of \tool in recommending arguments which are variables (parameters, local variables, class/instance variables). This kind of arguments accounts for 54.23\% of all the arguments. We found that for local variables and parameters\footnote{The arguments which are local variables and parameters are parts of \textit{Simple Name} in Table~\ref{tab:top_1_expr_type}}, \tool's top-1 accuracy is 88.54\%. Meanwhile, for others such as class variables and instance variables, the corresponding figure is 64.28\%. Although local variables and parameters usually are method-specific and less frequent, they are frequently used as arguments. This property of argument usage is captured by \tool thanks to features \textit{parameter similarity}, \textit{creating-distance}, and \textit{accessing-recentness}.

\begin{table}[]
\small
\centering
\caption{Impact of Context Length on Performance}
\label{tab:context_length}
\resizebox{\columnwidth}{!}{%
\begin{tabular}{lrrrrr}
\toprule
 & $l_1$ & $l_2$ & $l_3$ & $l_4$ & $l_5$ \\
 \midrule
Top-1 (\%)              & 62.05 & 65.86 & 66.14 & 67.00 & 67.83\\
\textit{MRR}                     & 0.70 & 0.72 & 0.72 & 0.73 & 0.74 \\
Run. time (s)        & 0.33 & 0.39 & 0.42 & 0.51 & 0.56 \\
\bottomrule
\end{tabular}
}
\end{table}


\subsubsection{Impact of testing context length}
Because the LMs in \tool recommend based on the given code sequences (the containing classes), the length of the code sequence could impact \tool's accuracy. To measure that impact, we randomly picked a fold of Netbeans for testing and the remaining folds for training in this experiment. 
For each AR request $r=\langle P, m\_call, pos \rangle$ whose containing class in context $P$ has $n$ code tokens, $C=\langle t_1, t_2, ..., t_n \rangle$, we evaluated the performance of \tool in five different context lengths. 
To do this, we varied the context-beginning point in five locations, $l_1$ -- $l_5$: 
the $i^{th}$ point $l_i$ contains $i \times \lfloor n/5 \rfloor$ last tokens of $C$. 
Formally, the code sequence fed to LMs is $C_i = \langle t_k, t_{k+1}, ..., t_n \rangle$ where $k=n - (i \times \lfloor n/5 \rfloor)$. In Table \ref{tab:context_length}, both top-1 accuracy and \textit{MRR} increase by 9.3\% and 5.7\%, respectively, when we move the point from $l_1$ to $l_5$, while running time increases 1.5 times. This is expected because \tool has more information to decide the likelihood of candidates then performs better. However, \tool needs more time to process longer contexts.

\subsubsection{Impact of origin of calling method}
For \textit{intra-project} method calls (49\% of the AR requests) where calling methods are declared in the requesting projects, calling methods might not be frequently used and are less likely to be seen by the models in the training set. Meanwhile, the \textit{inter-project} method calls (51\% of the AR requests) are usually the methods in common libraries such as JDK, Apache libraries, and Google Guava, and appear frequently in other projects. In this work, we also studied the impact of calling method's origin on \tool's performance. Particularly, \tool performed quite stably for \textit{intra-project} and \textit{inter-project} method calls, 72.9\% and 63.8\% top-1 accuracy, respectively. 
For \textit{intra-project} calls, the impact of less frequent calls is mitigated thanks to valid candidate identification and static features. Meanwhile, as the methods, which are designed to be used cross-project, they often accept quite general argument data types such as \texttt{String} or \texttt{Object}. The identified sets of valid candidates for \textit{inter-project} calls could be very large and contain many unexpected candidates (much \textit{noise}). 
On the other hand, the sets of valid candidates for \textit{intra-project} calls are much smaller because they require project-specific data types. This causes the higher accuracy for \textit{intra-project} calls compared to \tool's accuracy for \textit{inter-project} calls.

\subsubsection{Impact of training data size}
\begin{figure}
    \centering
    \includegraphics[width=1\columnwidth]{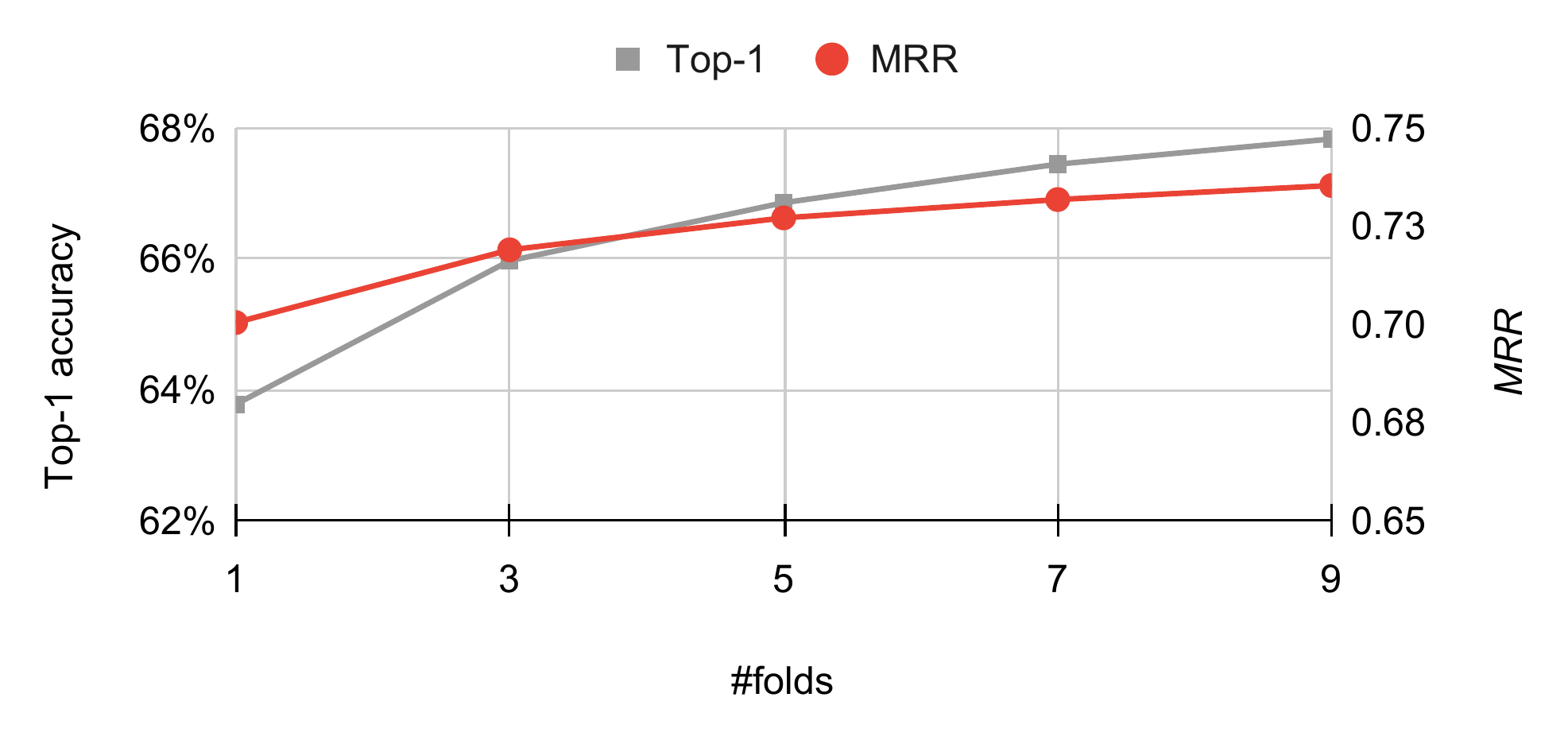}
    \caption{Impact of training data size}
    \label{fig:acc_folds}
\end{figure}

To study the impact of training data size on \tool's performance, we randomly picked a fold of Netbean for testing and varied the training data size by gradually adding more data by fold to the training data. As expected, the accuracy slightly improves when we increase the training data size (Figure \ref{fig:acc_folds}) because the models have observed more and thus performed better. In particular, the top-1 accuracy and \textit{MRR} increase by more than 3.4\% and 2.6\%, when data increases from 1 fold to 3 folds. However, the increasing trends slow down when data size increases from 3 folds to 9 folds. The reason is that the added data does not contain much more new knowledge as in the smaller sizes.

\subsection{Intrinsic Evaluation Results (RQ7)}

\subsubsection{Impact of Valid Candidate Identification}
To study the contribution of the valid candidate identification to the performance, we built a variant of \tool where the candidate identification is disabled and evaluated it on the large corpus and \textit{working-project} scenario.
In this variant, we applied a beam search algorithm to construct the whole arguments from tokens recommended by the light-ranking model. 
As seen in Table \ref{tab:intrinsic_candidate_gen}, the valid candidate identification component significantly contributes to both the recommendation accuracy and running time of \tool. Particularly, without the candidate identification, the performance decreases by 30\% for top-1 accuracy and 50\% for \textit{MRR}, while the recommendation process is two times slower than the full version of \tool.
As expected, we found that in most of the cases where the full version worked better, the $n$-gram LM with nested cache constructed many invalid candidates. This confirms the effectiveness of our strategy to identify valid candidates before ranking.

Furthermore, we measured the performance of the candidate identification in \tool. There are 91.0\% of the requests where the expected arguments are correctly identified. The average number of identified candidates for a request (before candidate reduction) is about 7.2K. 

 

\begin{table}[]
\small
   
      \caption{Impact of Valid Candidate Identification}
      \centering
      \label{tab:intrinsic_candidate_gen}
        \begin{tabular}{lccc}
            \toprule
             & \multicolumn{1}{c}{Top-1 (\%)} & \multicolumn{1}{c}{\textit{MRR}} & \multicolumn{1}{c}{Run. time (s)} \\
             \midrule
             
            ON & 69.96 & 0.76 & 0.444 \\
            OFF & 47.50 & 0.51 & 0.809 \\
        \bottomrule
        \end{tabular}%
    
\end{table}

\begin{table}[]
\small

\centering
\caption{Impact of Candidate Reduction}
\label{tab:intrinsic_cand_reduction}
\begin{tabular}{lccc}
\toprule
 & \multicolumn{1}{c}{Top-1 (\%)} & \multicolumn{1}{c}{\textit{MRR}} & \multicolumn{1}{c}{Run. time (s)} \\
 \midrule
 
ON & 69.96 & 0.76 & 0.444 \\
OFF & 61.98 & 0.69 & 2.424 \\
\bottomrule
\end{tabular}%
    
\end{table}

\subsubsection{Impact of Candidate Reduction}
We constructed a variant of \tool without the candidate reduction component to measure the component's contribution to the performance of \tool with our large corpus and \textit{working-project} scenario (Table \ref{tab:intrinsic_cand_reduction}). With much larger candidate sets (+7K candidates on average), \tool's performance is greatly reduced in both accuracy and efficiency when disabling the candidate reduction component. Specially, both top-1 accuracy and \textit{MRR} significantly decrease, by about 11\% and 8\%, respectively. 
Moreover, the variant is also much slower, 6 times (2.4 sec) slower than that when enabling this component. 
%
%
%
Thus, candidate reduction should be enabled to maintain both the performance of \tool.

\subsubsection{Impact of Light-ranking Model}
To evaluate the impact of selecting the light-ranking model, we built two variants of \tool with different count-based LMs applied for its light-ranking component: (plain) $n$-gram LM~\cite{n_gram} and $n$-gram LM with nested-cache~\cite{nestedcache}. In both variants, $n=6$ and Jelinek-Mercer smoothing are applied, the reduction threshold is 20, and GPT-2 is applied in the heavy-ranking stage. For the large corpus, our results show that \tool with plain $n$-gram LM achieved top-1 accuracy of only about 53\%, while the corresponding figure of \tool when applying $n$-gram LM with nested-cache was about 70\% with a marginal running time increase. These results are expected and consistent with the findings of Hellendoorn \etal~\cite{nestedcache} where the $n$-gram LM carefully adapted for source code can yield performance that surpasses the plain $n$-gram LM.

\subsubsection{Impact of Reduction Threshold}
\label{sec:rt-impact}
Additionally, we evaluated the reduction threshold, $RT$, on the accuracy and running time of \tool. We varied $RT$ from 10 to 50 to study its impact on \tool's performance (Table \ref{tab:threshold}).
As seen, when $RT=10$, \tool ran faster because of small reduced sets of valid candidates. However, with a small reduction threshold, in many requests, the expected arguments were incorrectly eliminated before feeding the reduced candidate set to the heavy-ranking stage. The accuracy of \tool is slightly improved when increasing $RT$ from $10$ to $50$. This is because with a larger threshold, there are fewer requests where the identified candidate sets are over-reduced. Meanwhile, the running time is significantly increased because the heavy-ranking stage accounts for about 86.3\% of the total running time (Section \ref{sec:time_complexity}). 
Thus, in this version, we kept $RT = 20$ to balance between the accuracy and recommending time.


\begin{table}[!htb]

\centering
\caption{Impact of reducing threshold, $RT$}
\label{tab:threshold}
\resizebox{\columnwidth}{!}{%
\begin{tabular}{lccccc}
\toprule
$RT$ & \multicolumn{1}{r}{10} & \multicolumn{1}{r}{20} & \multicolumn{1}{r}{30} & \multicolumn{1}{r}{40} & \multicolumn{1}{r}{50}  \\
\midrule
Top-1 (\%)          & 63.77 & 64.67 & 65.10 & 65.34 & 65.49 \\
\begin{tabular}[c]{@{}l@{}}Run.\\ time (s)\end{tabular}    & 0.342 & 0.406 & 0.418 & 0.464 & 0.508 \\
\toprule
\end{tabular}%
}
\end{table}

\begin{table}[!htb]

\centering
\caption{Impact of heavy-ranking stage}
\label{tab:global-model}
\begin{tabular}{lccc}
\toprule
$\mathcal{P}_{hr}$ & \multicolumn{1}{c}{Top-1 (\%)} & \multicolumn{1}{c}{\textit{MRR}} & \multicolumn{1}{c}{Run. Time (s)} \\
 \midrule
OFF                 & 65.37 & 0.72 & 0.090 \\
GPT-2               & 70.71 & 0.76 & 0.732 \\
CodeT5              & 68.59 & 0.74 & 0.186 \\
LSTM                & 49.26 & 0.61 & 0.198 \\
$n$-gram            & 36.89 & 0.51 & 0.137 \\
\bottomrule
\end{tabular}%
   
\end{table}

\subsubsection{Impact of Candidate Ranking Component}
We also studied the impact of selecting heavy-ranking model on the overall performance of \tool. In \tool's process, any approach that can evaluate the likelihood of candidates could be applied as a heavy-ranking model. In this work, we selected several representative approaches for the evaluation. In this experiment, we randomly selected a fold of Netbeans for testing and the remaining folds for training to evaluate all variants of \tool.
In Table \ref{tab:global-model}, \tool without heavy-ranking stage (OFF) can run very fast (0.09s), yet achieved about 65.37\% of top-1 accuracy. The top-1 accuracy of \tool can be improved by 5\% and 8.2\% when applying CodeT5 and GPT-2, respectively. This is expected because GPT-2 and CodeT5 can capture dependencies which are longer than that by the light-ranking model. However, the running time when using these powerful models greatly increases by about 8.6 times when GPT-2 is applied. With the less powerful LMs such as vanilla LSTM and $n$-gram, the performance of \tool might not be improved, or even be significantly worsened.

\subsubsection{Impact of Static Features}
To investigate the impact of the static features used in this version of \tool: \textit{parameter similarity}, \textit{creating-distance}, and \textit{accessing-recentness}, we built a variant of \tool where the application of only these static features is disabled. 
In this experiment, Eclipse is used for both \tool and the variant.
We found that the full version of \tool's top-$1$ accuracy is 0.4\% better than that of the variant where the static features are not applied. The reason for this marginal improvement is that the heavy-ranking model applied in the current version of \tool is GPT-2. This powerful model can dynamically learn these features to rank candidates. Still, for LSTM and $n$-gram which are less powerful, these features can improve 45.4\% and 20\% respectively in top-1 accuracy. Thus, these static features should be enabled to ensure \tool's best performance for any models used in \tool.

\subsection{Time Complexity (RQ8)}
\label{sec:time_complexity}
All experiments were run on a server with Intel(R) Xeon(R) CPU @ 2.30GHz, 32GB RAM, and Tesla P100 GPU. 
\tool took 38 hours for training. Our average recommendation time is 0.444s/request.
Valid candidate identifying took 9\%, while candidate reducing and ranking took 4.7\% and 86.3\%. 

Analyzing the impact of \tool's components on its efficiency, we found that the valid candidate identification (VCI) step significantly affects \tool's efficiency. As seen in Table~\ref{tab:intrinsic_candidate_gen}, \tool with VCI recommended almost 2 times faster than when the beam search technique is applied.
Meanwhile, Candidate Reduction reduces \tool's running time by about 6 times compared to that when that component is disabled (Table~\ref{tab:intrinsic_cand_reduction}). 
%
%
This is because with only less than 5\% of overall running time of \tool, the candidate reducing step significantly narrows down the candidate sets, from 7K to less than 50.
With those much smaller sets of candidates, the ranking stage, which accounts for 86.3\% the overall \tool's recommending time, saves much time. As a result, \tool's overall running time of  is greatly reduced.
 Additionally, the \tool's running time is also proportional to the reducing threshold (Table~\ref{tab:threshold}). 
Table~\ref{tab:global-model} shows that selecting heavy-ranking models could greatly impact \tool's efficiency. For example, \tool  with GPT-2 is 8 times slower than \tool without the heavy-ranking stage. 

\subsection{Threats to Validity}
\label{sec:threats}
The main threats to the validity of our work consist of internal, construct, and external threats.

\textbf{Threats to internal validity} include the influence of the hyper-parameters used in \tool. \tool's performance would be affected by different hyper-parameter settings, which are tuned empirically in our empirical study. 
Thus, there is a threat to the hyper-parameter choosing, and further improvement is possible. However, current settings have achieved acceptable performance. 
Another threat to validity relates to using the beam search algorithm in our adaptation of the baseline methods for recommending arguments. 
To reduce this threat and balance the cost of a practical evaluation, we used a relatively large beam size for the beam search algorithm.
%

\textbf{Threats to construct validity} relate to the suitability of our evaluation procedure. We used \textit{Precision}, \textit{Recall}, top-$k$ accuracy, and \textit{MRR}. They are the classical evaluation measures for argument recommendation and code completion and are used in almost all the previous AR and code completion work~\cite{PARC,Precise,bigcode,ase19,curglm}. In fact, a user study could be the best evaluation method, which investigates how developers interact with the proposed approach and whether the proposed approach can indeed help developers find their expected arguments. In future work, we are planning to perform a human evaluation to measure actual performance of \tool in a real-world setting.
%

\textbf{Threats to external validity} mainly lie in the procedure used in our experiments. Our data has only Java code. Thus, we cannot claim that similar results would have been observed in other programming languages, especially for dynamically typed languages (untyped languages). Further studies are needed to validate and generalize our findings to other languages. Another threat relates to the quality of the datasets. To reduce this threat, we used the same very large datasets which are widely used in existing studies. A threat relates to our selections for baseline approaches. To reduce this threat, we selected the state-of-the-art $n$-gram with nested cache, CodeT5~\cite{codet5}, and the combination of GPT-2~\cite{gpt2} and a type-based post-processing step. Moreover, $n$-gram with nested cache is used to represent for the classical direction which is widely used in existing studies~\cite{bigcode,allamanis2018survey,failures}. Meanwhile, CodeT5 and GPT-2 are used as a representative advanced DNN approach instead of vanilla LSTM or RNN. Moreover, GPT-2 is used as the core engine of Tabnine~\cite{tabnine} and IntelliCode~\cite{intellicode} which are two of the most popular AI-assisted code completers.

\section{Related Work}
\label{sec:related_work}

\textbf{Argument recommendation}. \tool is related to PARC by Asaduzzaman \etal~\cite{PARC} and Precise by Zhang \etal~\cite{Precise}. In comparison, there are fundamental differences between \tool and both PARC and Precise. First, PARC and Precise follow IR direction which searches for similar arguments that have been seen in the corpus. The seen arguments which are represented by a set of static features are stored in a database. Similar arguments are found by \textit{simhash} algorithm in PARC~\cite{PARC} and $k$-nearest neighbor algorithm in Precise~\cite{Precise}. Meanwhile, \tool is a learning-based approach applying statistical language models learning the dynamic features to recommend arguments. 
Secondly, in the recommendation process, the syntactical and semantic constraints are enforced very differently in PARC and Precise as \tool. In PARC and Precise, the type-compatibility constraint is used to validate the similar argument candidates at the last step. 
The validation becomes not very useful when the expected arguments are unseen or not sufficiently similar to the seen arguments. For \tool, these constraints are applied to identify the set of valid candidates for the given request at the first step. Our results show that \tool is very effective in identifying the valid argument candidates, i.e., the expected arguments are identified in +90\% of the requests. This ensures that ranking models work in the valid set of candidates even for unseen expected arguments.
The empirical results also show \tool's ability to recommend unseen arguments.

\textbf{API recommendation}. There exists a rich literature of approaches on API recommendation. These studies mainly focus on suggesting methods to call. In modern IDEs such as Eclipse~\cite{eclipse_rec} or Intellij 
IDEA~\cite{intellij_rec}, APIs are suggested primarily based on program analysis, which utilizes type-compatibility and accessibility information to support completion of method calls/arguments in the current context. 
To improve the API recommendation performance, instead of just listing all valid candidates alphabetically, several approaches also leverage previous API usages to rank, filter, or group candidates for better suggestions~\cite{bruch2009learning, hou2010towards,hou2011evaluation,rahman2016rack,gu2016deep,nguyen2016api,huang2018api,chen2021api}. 
%
%
Huang \etal~\cite{huang2018api} discover task-API knowledge gap and the necessity of incorporating API documentation and Stack Overflow posts to search for relevant API usages. 
Recently, Chen \etal~\cite{chen2021api} combine the textual and dependency information of source code, which analyzes control and data flow graphs to capture correlated API usages and overcome the limitation on long dependencies of token-sequence-based API recommendation. Kang \etal~\cite{kang2021apirecx} apply pre-trained LM to recommend Cross-Library APIs. 
However, while those approaches focus on recommending method calls, \tool is designed to predict actual parameters for method calls. Thus, \tool and those approaches can be applied together to recommend longer code sequences.



\textbf{Code completion and suggestion}. Traditional work on code completion mainly focused on defining heuristic rules based on language-specific grammar and type-checking information to complete current code fragments~\cite{hou2010towards, hou2011evaluation}. Hindle \etal~\cite{naturalness} show that source code is highly repetitive (\textit{natural}). Several studies~\cite{naturalness, tu2014localness,nguyen2013statistical,nestedcache,ase19,allamanis2018survey} have been proposed, exploiting this naturalness property to predict next tokens. Tu \etal~\cite{tu2014localness} demonstrate that code also has localness property, which could be exploited by the combination of cache mechanism and traditional $n$-gram model. Hellendoorn and Devanbu~\cite{nestedcache} introduce nested-cache, an $n$-gram model integrated with nested scopes, locality, and dynamism, which showed promising performance on code completion task. 

In recent years, deep neural network (DNN) language models have made great progress while applying to learn source code~\cite{bigcode,li2017code, curglm,liu2016neural,he2021pyart,codefill}. Neural networks such as RNN and vanilla LSTM could capture longer dependencies compared to $n$-gram model, which could benefit code completer in large context circumstances~\cite{liu2016neural}. In fact, training a DNN language model requires large datasets, which might produce a huge vocabulary set and cause Out-of-Vocabulary (OOV) problem~\cite{li2017code, bigcode}. Li \etal ~\cite{li2017code} introduce a pointer mixture model to mitigate the OOV issue. Karampatsis \etal~\cite{bigcode} propose a solution for large-scale open-vocabulary DNN language models, which employ the BPE algorithm, beam search algorithm, and cache mechanism. 
%
%
%

Although these code completion approaches are originally designed to predict any next tokens and not whole arguments, they can be adapted for argument recommendation (e.g., applying beam search algorithm). 
We showed that \tool outperforms the representative code completion approaches. In fact, any existing model for code completion can be applied as a ranking model in our argument recommendation approach.



\textbf{Learning-based approaches for SE tasks}. Several studies have been proposed for specific SE tasks including code suggestion~\cite{icse20, naturalness, allamanis2016convolutional}, program synthesis~\cite{amodio2017neural,gvero2015synthesizing}, pull request description generation~\cite{hu2018deep,liu2019automatic}, code summarization~\cite{iyer2016summarizing,mastropaolo2021studying,wan2018improving}, code clones~\cite{li2017cclearner}, fuzz testing\cite{godefroid2017learn}, fault localization~\cite{li2021fault}, and program repair~\cite{jiang2021cure,ding2020patching}.

\section{Conclusion}
\label{sec:conclusion}
We introduce \tool, a novel automated argument recommendation approach combining program analysis (PA) and statistical language models (LMs) in the recommendation process. In \tool, the statistical LMs are used to suggest the frequent/natural candidates identified by applying PA. Meanwhile, PA is applied to navigate LMs working on the set of valid candidates. Our empirical evaluation on the large datasets of real-world projects shows that \tool improves the state-of-the-art approach up to 19\% and 18\% in both precision and recall for recommending arguments of \textit{frequently-used libraries}. For \textit{general argument recommendation task}, \tool also outperforms the baseline approaches by up to 31\% and 125\% top-1 accuracy. Moreover, for new projects, \tool achieves more than 60\% top-3 accuracy after training with a larger dataset. Especially, with solely personalizing the light-ranking stage which is deployed on developers' machines, \tool can correctly rank the expected arguments at the top-1 position in up to +7/10 recommendation requests in less than 0.5 seconds while preserving the privacy of developers.

\printcredits
\bibliographystyle{elsarticle-num}

\bibliography{9.references}

\end{document}